\documentclass[iop,twocolappendix]{emulateapj}

\newcommand{\Kepler}{{\it Kepler}}

\usepackage{color}

\usepackage[export]{adjustbox}

\usepackage{makecell}
\usepackage{hhline}

\usepackage{amsmath}
\usepackage{longtable}
\usepackage{afterpage}

\usepackage{wasysym}
\usepackage{soul}

\shorttitle{K2-136\MakeLowercase{c}}
\shortauthors{Mayo et al.}

\begin{document}


\title{Hyades Member K2-136\MakeLowercase{c}:\\The Smallest Planet in an Open Cluster with a Precisely Measured Mass}


\author{Andrew W. Mayo\altaffilmark{1,2,$\dagger$,$\ddagger$}, Courtney D. Dressing\altaffilmark{1}, Andrew Vanderburg\altaffilmark{3}, Charles D. Fortenbach\altaffilmark{4}, Florian Lienhard\altaffilmark{5}, Luca Malavolta\altaffilmark{7}, Annelies Mortier\altaffilmark{5,6,8}, Alejandro N\'u\~nez\altaffilmark{10}, Tyler Richey-Yowell\altaffilmark{11}, Emma V. Turtelboom\altaffilmark{1}, Aldo S. Bonomo\altaffilmark{12}, David W. Latham\altaffilmark{13}, Mercedes L\'opez-Morales\altaffilmark{13}, Evgenya Shkolnik\altaffilmark{11}, Alessandro Sozzetti\altaffilmark{12}, Marcel A. Ag\"ueros\altaffilmark{9,10}, Luca Borsato\altaffilmark{14}, David Charbonneau\altaffilmark{13}, Rosario Cosentino\altaffilmark{15}, Stephanie T. Douglas\altaffilmark{16}, Xavier Dumusque\altaffilmark{17}, Adriano Ghedina\altaffilmark{15}, Rose Gibson\altaffilmark{18}, Valentina Granata\altaffilmark{7,14}, Avet Harutyunyan\altaffilmark{15}, R. D. Haywood\altaffilmark{19}, Gaia Lacedelli\altaffilmark{7,14}, Vania Lorenzi\altaffilmark{15,20}, Antonio Magazz\`u\altaffilmark{15}, A. F. Martinez Fiorenzano\altaffilmark{15}, Giuseppina Micela\altaffilmark{21}, Emilio Molinari\altaffilmark{22}, Marco Montalto\altaffilmark{7}, Domenico Nardiello\altaffilmark{23,14}, Valerio Nascimbeni\altaffilmark{14}, Isabella Pagano\altaffilmark{24}, Giampaolo Piotto\altaffilmark{7,14}, Lorenzo Pino\altaffilmark{25}, Ennio Poretti\altaffilmark{15,26}, Gaetano Scandariato\altaffilmark{23}, Stephane Udry\altaffilmark{16}, Lars A. Buchhave\altaffilmark{2,27}}

\altaffiltext{$\dagger$}{\texttt{mayo@berkeley.edu}}
\altaffiltext{$\ddagger$}{National Science Foundation Graduate Research Fellow}
\altaffiltext{1}{Astronomy Department, University of California, Berkeley, CA 94720, USA}
\altaffiltext{2}{Centre for Star and Planet Formation, Natural History Museum of Denmark \& Niels Bohr Institute, University of Copenhagen, \O ster Voldgade 5-7, DK-1350 Copenhagen K., Denmark}
\altaffiltext{3}{Department of Physics and Kavli Institute for Astrophysics and Space Research, Massachusetts Institute of Technology, 77 Massachusetts Avenue, Cambridge, MA,
02139, USA}
\altaffiltext{4}{Department of Physics and Astronomy, San Francisco State University, San Francisco, CA 94132, USA}
\altaffiltext{5}{Astrophysics Group, Cavendish Laboratory, University of Cambridge, J. J. Thomson Avenue, Cambridge CB3 0HE, UK}
\altaffiltext{6}{School of Physics \& Astronomy, University of Birmingham, Edgbaston, Birmingham, B15 2TT, UK}
\altaffiltext{7}{Dipartimento di Fisica e Astronomia Galileo Galilei, Universitá di Padova, Vicolo dell'Osservatorio 3, 35122 Padova, Italy}
\altaffiltext{8}{Kavli Institute for Cosmology, University of Cambridge, Madingley Road, Cambridge CB3 0HA, UK}
\altaffiltext{9}{Laboratoire d'astrophysique de Bordeaux, Univ.~Bordeaux, CNRS, B18N, All\'ee Geoffroy Saint-Hilaire, 33615 Pessac, France}
\altaffiltext{10}{Department of Astronomy, Columbia University, 550 West 120th Street, New York, NY 10027, USA}
\altaffiltext{11}{School of Earth and Space Exploration, Arizona State University, Tempe, AZ 85287, USA}
\altaffiltext{12}{INAF – Osservatorio Astrofisico di Torino, Via Osservatorio 20, 10025 Pino Torinese, Italy}
\altaffiltext{13}{Center for Astrophysics ${\rm \mid}$ Harvard {\rm \&} Smithsonian, 60 Garden Street, Cambridge, MA 02138, USA}
\altaffiltext{14}{INAF - Osservatorio Astronomico di Padova, Vicolo dell'Osservatorio 5, IT-35122, Padova, Italy}
\altaffiltext{15}{Fundaci\'on Galileo Galilei-INAF, Rambla Jos\'e Ana Fernandez P\'erez 7, 38712 Bre\~na Baja, Tenerife, Spain}
\altaffiltext{16}{Department of Physics, Lafayette College, 730 High Street, Easton, PA 18042, USA}
\altaffiltext{17}{Department of Astronomy of the University of Geneva, Geneva Observatory, Chemin Pegasi 51, 1290 Versoix, Switzerland}
\altaffiltext{18}{Columbia University, Department of Astronomy, 550 West 120th Street, New York, USA, 10027}
\altaffiltext{19}{Astrophysics Group, University of Exeter, Exeter EX4 2QL, UK}
\altaffiltext{20}{Instituto de Astrofísica de Canarias, C/Vía Láctea s/n, 38205 La Laguna, Spain}
\altaffiltext{21}{INAF - Osservatorio Astronomico di Palermo, Piazza del Parlamento 1, I-90134 Palermo, Italy}
\altaffiltext{22}{INAF – Osservatorio Astronomico di Cagliari \& REM, Via della Scienza, 5, 09047 Selargius CA, Italy}
\altaffiltext{23}{Aix Marseille Univ, CNRS, CNES, LAM, Marseille, France}
\altaffiltext{24}{INAF – Osservatorio Astrofisico di Catania, Via Santa Sofia 78, 95123 Catania, Italy}
\altaffiltext{25}{INAF - Osservatorio Astrofisico di Arcetri, Largo Enrico Fermi 5, 50125, Firenze, Italy}
\altaffiltext{26}{INAF – Osservatorio Astronomico di Brera, Via E. Bianchi 46, 23807 Merate, Italy}
\altaffiltext{27}{DTU Space, National Space Institute, Technical University of Denmark, Elektrovej 327, DK-2800 Lyngby, Denmark}


\begin{abstract}
K2-136 is a late-K dwarf ($0.742\pm0.039$ M$_\odot$) in the Hyades open cluster with three known, transiting planets and an age of $650\pm70$ Myr. Analyzing \emph{K2} photometry, we found that planets K2-136b, c, and d have periods of $8.0$, $17.3$, and $25.6$ days and radii of $1.014\pm0.050$ R$_\oplus$, $3.00\pm0.13$ R$_\oplus$, and $1.565\pm0.077$ R$_\oplus$, respectively. We collected 93 radial velocity measurements (RVs) with the HARPS-N spectrograph (TNG) and 22 RVs with the ESPRESSO spectrograph (VLT). Analyzing HARPS-N and ESPRESSO data jointly, we found K2-136c induced a semi-amplitude of $5.49\pm0.53$ m s$^{-1}$, corresponding to a mass of $18.1\pm1.9$ M$_\oplus$. We also placed $95\%$ upper mass limits on K2-136b and d of $4.3$ and $3.0$ M$_\oplus$, respectively. Further, we analyzed HST and XMM-Newton observations to establish the planetary high-energy environment and investigate possible atmospheric loss. K2-136c is now the smallest planet to have a measured mass in an open cluster and one of the youngest planets ever with a mass measurement.
K2-136c has $\sim$75\% the radius of Neptune but is similar in mass, yielding a density of $3.69^{+0.67}_{-0.56}$ g cm$^{-3}$ ($\sim$2-3 times denser than Neptune). Mass estimates for K2-136b (and possibly d) may be feasible with more RV observations, and insights into all three planets' atmospheres through transmission spectroscopy would be challenging but potentially fruitful. This research and future mass measurements of young planets are critical for investigating the compositions and characteristics of small exoplanets at very early stages of their lives and providing insights into how exoplanets evolve with time.
\end{abstract}

\keywords{planets and satellites: composition - planets and satellites: detection - planets and satellites: fundamental parameters - planets and satellites: rocky planets - methods: data analysis - techniques: radial velocities \vspace{-5mm}
}

\section{Introduction} \label{intro}

The timescales on which planets and planetary systems evolve are far longer than any feasible timescale of scientific observations. The only way to learn about how planets form and evolve is to collect snapshots at different stages of their development and assemble these snapshots into a cohesive framework. This is where open clusters prove particularly useful. Open clusters, close collections of young, recently formed stars, are excellent laboratories for studying the early lives of stars, because all of the stars in an open cluster, regardless of size, temperature, metallicity, or location, have a shared formation history, and therefore the ages of the stars can be very tightly constrained. This logic can also be applied to planets; if they form very quickly after the coalescence of their host star \citep{raymonetandmorbidelli2020}, it is possible to determine the age of a planet orbiting an open cluster star, thereby capturing one of the early snapshots required to assemble the framework of a planet's evolution.

In this paper we characterize K2-136c, a sub-Neptune planet in the Hyades open cluster. Orbiting a late K dwarf, this planet is one of the three known, transiting planets in the system. The system was originally observed in \emph{K2} Campaign 13 for 80 days (2017 March 8 - 2017 May 27) and was proposed for observation by seven guest observer teams: GO13008, GO13018, GO13023, GO13049, GO13064, GO13077, and GO13090. All three planets were originally discovered by \citet{mannetal2018} (hereafter M18) and \citet{ciardietal2018} (a parallel analysis published simultaneously). Shortly thereafter a subsequent analysis was completed by \citet{livingstonetal2018}. All three papers are in broad agreement regarding stellar and planetary parameters, but M18 established the tightest constraints on orbital period for all three planets.

In their analysis, M18 found an Earth-sized planet ($0.99^{+0.06}_{-0.04}$ R$_\oplus$) at $P = 8.0$ days (K2-136b), a sub-Neptune-sized planet ($2.91^{+0.11}_{-0.10}$ R$_\oplus$) at $P = 17.3$ days (K2-136c, the focus of this paper), and a super-Earth-sized planet ($1.45^{+0.11}_{-0.08}$ R$_\oplus$) at $P = 25.6$ days (K2-136d). They also determined a host star mass of $0.74 \pm 0.02$ M$_\odot$ and a stellar radius of $0.66 \pm 0.02$ R$_\odot$.

As for the stellar age, there are a number of estimates available. \citet{perrymanetal1998} found the Hyades open cluster to be $625 \pm 50$ Myr. \citet{gossageetal2018} found an age of $\sim 680$ Myr while \citet{brandtandhuang2015} determined a slightly older age of $750 \pm 100$ Myr. The age we use throughout this paper comes from \citet{martinetal2018}, who determined the Hyades to be $650 \pm 70$ Myr old. We thus assume that K2-136 and the three orbiting planets share that approximate age. We chose this age because it is a relatively recent result, it compares and combines results using both old \citep{burrowsetal1997} and new \citep{baraffeetal2015} standard evolutionary models, and it also agrees broadly with other, previous estimates. The young age of the system was our primary reason for pursuing K2-136c as a target: there are very few young, small planets with mass measurements. According to the NASA Exoplanet Archive (accessed 2023 Mar 12; \citealt{https://doi.org/10.26133/nea12}), there are only 13 confirmed exoplanets with $R_p < 4$ R$_\oplus$, a host star age $< 1$ Gyr, and a mass measurement (not an upper limit): HD 18599b \citep{desideraetal2022}, HD 73583b and c \citep{barraganetal2022}; K2-25b \citep{stefanssonetal2020}; L 98-59b, c, and d \citep{demangeonetal2021}; Kepler-411b and Kepler-411d \citep{sunetal2019}; Kepler-462b \citep{masudaetal2020}; Kepler-289b and Kepler-289d \citep{schmittetal2014}; and K2-100b \citep{barraganetal2019}. Of these, only the Kepler-411, K2-100, HD 73583, K2-25, and HD 18599 systems have an age constraint tighter than $50\%$ \citep{sunetal2019,barraganetal2019,barraganetal2022,stefanssonetal2020,desideraetal2022}.

We analyzed photometry of the K2-136 system in order to measure the radii, ephemerides, and other transit parameters of each planet. We also collected spectra of the K2-136 system and measured radial velocities (RVs) as well as stellar activity indices. Then, by modeling these RVs (following \citealt{rajpauletal2015}), we determined the mass of K2-136c and placed upper limits on the masses of the other two planets. We used this system to investigate the nature, environment, and evolution of young, small exoplanets.

This paper is organized as follows. In Section~\ref{obs} we discuss our observations. Then we detail our method of stellar characterization in Section~\ref{stellar_char}. Next, in Section~\ref{analysis} we describe our RV and photometry models, data analysis, model comparison, and parameter estimation. In Section~\ref{results_discussion} we present and discuss our results. Finally, we summarize and conclude in Section~\ref{conclusion}.

\section{Observations} \label{obs}

\subsection{\emph{K2}}

Photometric observations of the K2-136 system were collected with the \Kepler\ spacecraft \citep{boruckietal2008} through the \emph{K2} mission during Campaign 13 (2017 Mar 08 to 2017 May 27). \emph{K2} collected long-cadence observations of this system every $29.4$ minutes.

\subsection{\emph{TESS}}

Photometric observations of the K2-136 system were also collected with the \emph{TESS} spacecraft \citep{rickeretal2015} during Sector 43 (2021 Sep 16 to 2021 Oct 10) and Sector 44 (2021 Oct 12 to 2021 Nov 06). \emph{TESS} collected long-cadence full frame image observations of this system every $10$ minutes in Sector 43 and short-cadence observations every $20$ seconds in Sector 44.

\subsection{HARPS-N}

We collected 93 RV observations using the HARPS-N spectrograph (\citealt{cosentinoetal2012}, \citealt{cosentinoetal2014}) on the Telescopio Nazionale Galileo (TNG). The first 88 spectra were collected between 2018 August 11 and 2019 February 7 (programs A37TAC\_24 and A38TAC\_27, PI: Mayo), and the final 5 spectra were collected between 2020 September 18 and 2020 October 31 by the HARPS-N Guaranteed Time Observation program. RVs and additional stellar activity indices were extracted using a K6 stellar mask and version 2.2.8 of the Data Reduction Software (DRS) adapted from the ESPRESSO pipeline. Spectra had an average exposure time of $1776.5$ seconds and the average SNR in the order around $550$ nm was $51.1$. The RV standard deviation was $6.9$ m s$^{-1}$ and the RV median uncertainty was $1.6$ m s$^{-1}$. Stellar activity indices also extracted and reported in this paper include the cross-correlation function (CCF) bisector span inverse slope (hereafter BIS), the CCF full width at half maximum (FWHM), and $S_{HK}$ (which measures chromospheric activity via core emission in the Ca II H and K absorption lines). The observation dates, velocities, and activity indices are provided in Table~\ref{table:rv_table}. 

\subsection{ESPRESSO}

We collected 22 RV observations using the ESPRESSO spectrograph \citep{pepeetal2021} on the Very Large Telescope (VLT) between 2019 November 1 and 2020 February 27 (program 0104.C-0837(A), PI: Malavolta). RVs and additional stellar activity indices were extracted using a K6 stellar mask and the same pipeline as the HARPS-N observations (DRS version 2.2.8). Typical exposure time for spectra was $1800$ seconds and the average SNR at Order 111 (central wavelength = $551$nm) was $79.9$. The RV standard deviation was $7.8$ m s$^{-1}$ and the RV median uncertainty was $0.70$ m s$^{-1}$. These observations and indices are also provided in Table~\ref{table:rv_table}.

\subsection{Hubble Space Telescope} \label{HST_obs}

Near-ultraviolet (NUV) observations of K2-136 were taken as part of a broader {\it Hubble Space Telescope} ({\it HST}) program observing the Hyades (GO-15091, PI: Ag\"ueros). The target was exposed for 1166.88 seconds on 2019 September 13 using the photon-counting Cosmic Origins Spectrograph (COS; \citealt{greenetal2012}) in the G230L filter and had no data quality flags. 

After initial data reduction through the \texttt{CALCOS} pipeline version 3.3.10, we additionally confirmed that the star was not flaring during observations by integrating the background-subtracted flux by wavelength over 1 and 10 second time intervals in the time-tagged data. No flares above 3$\sigma$ were identified. 

\subsection{XMM-Newton} \label{XMM_obs}
K2-136 was the target of an {\it XMM-Newton} ({\it XMM}) 43~ksec observation on 2018 September 11 (Obs.~ID: 0824850201, PI: Wheatley). The observation was processed using the standard Pipeline Processing System (PPS version $17.56\_20190403\_1200$; Pipeline sequence ID: 147121). The source detection corresponding to K2-136 was detected by both the pn and MOS cameras, for a total of 800 source counts in the 0.2-12.0 keV energy band. The X-ray source has data quality flag \texttt{SUM\_FLAG}=0 (i.e., good quality). No variability or pileup were detected for this X-ray source.

\begin{table*}[t]
\begin{center}
\caption{Stellar parameters of K2-136\label{stellar_table}}
\begin{tabular}{llcccc}
\tableline
\tableline
Parameter & Unit & & & Value & Reference \\
\tableline\\
EPIC & - & .................. & .................. & 247589423 & - \\
2MASS & - & .................. & .................. & J04293897+2252579 & - \\
$\alpha$ R.A. & J2016.0 & .................. & .................. & 04:29:39.1 & GAIA DR3\footnote[1]{\citet{gaiacollaboration2016b,gaiacollaboration2020,babusiauxetal2022,gaiacollaboration2022}} \\
$\delta$ Dec & J2016.0 & .................. & .................. & +22:52:57.2 & GAIA DR3\footnotemark[1] \\
$\mu_{\alpha}$ & mas yr$^{-1}$ & .................. & .................. & $82.778 \pm 0.021$ & GAIA DR3\footnotemark[1] \\
$\mu_{\delta}$ & mas yr$^{-1}$ & .................. & .................. & $-35.541 \pm 0.015$ & GAIA DR3\footnotemark[1] \\
Parallax & mas & .................. & .................. & $16.982 \pm 0.019$ & GAIA DR3\footnotemark[1] \\
Distance & pc & .................. & .................. & $58.752^{+0.061}_{-0.072}$ & b\footnotetext[2]{\citet{bailer-jonesetal2021}} \\
Age & Myr & .................. & .................. & $650 \pm 70$ & c\footnotetext[3]{\citet{martinetal2018}} \\
B mag & - & .................. & .................. & $12.48 \pm 0.01$ & UCAC4\footnote[4]{\citet{zachariasetal2013}} \\
V mag & - & .................. & .................. & $11.20 \pm 0.01$ & UCAC4\footnotemark[4] \\
J mag & - & .................. & .................. & $9.096 \pm 0.022$ & 2MASS\footnote[5]{\citet{cutrietal2003,skrutskieetal2006}} \\
H mag & - & .................. & .................. & $8.496 \pm 0.020$ & 2MASS\footnotemark[5] \\
K mag & - & .................. & .................. & $8.368 \pm 0.019$ & 2MASS\footnotemark[5] \\
W1 mag & - & .................. & .................. & $8.263 \pm 0.023$ & WISE\footnote[6]{\citet{wrightetal2010}} \\
W2 mag & - & .................. & .................. & $8.349 \pm 0.020$ & WISE\footnotemark[6] \\
W3 mag & - & .................. & .................. & $8.312 \pm 0.030$ & WISE\footnotemark[6] \\
Fractional X-ray luminosity $L_\mathrm{X}/L_*$ & - & .................. & .................. & ($1.97 \pm 0.30$)$\times10^{-5}$ & This work \\
\tableline
\tableline
Parameter & Unit & SPC & ARES+MOOG & Combined\footnote[7]{Systematic uncertainties added in quadrature \citep{tayaretal2020}} & Reference \\
\tableline\\

Effective temperature $T_{\mathrm{eff}}$ & K & $4517 \pm 49$ & $4447 \pm 149$ & $4500^{+125}_{-75}$ & This work \\
Surface gravity $\log{g}$ & g cm$^{-2}$ & $4.68 \pm 0.10$ & $4.82 \pm 0.43$ & - & This work \\
Microturbulence & km s$^{-1}$ & - & $<1.1$\footnote[8]{Value is poorly constrained, $1\sigma$ upper limit reported instead \vspace{7mm}} & - & This work \\
Metallicity {[$\mathrm{Fe}/\mathrm{H}$]} & dex & - & $0.05 \pm 0.10$ & - & This work \\
Metallicity {[$\mathrm{M}/\mathrm{H}$]} & dex & $-0.02 \pm 0.08$ & - & - & This work \\
Radius $R_*$ & R$_\odot$ & $0.6764^{+0.0039}_{-0.0033}$ & $0.6770^{+0.0050}_{-0.0038}$ & $0.677 \pm 0.027$ & This work \\
Mass $M_*$ & M$_\odot$ & $0.7413^{+0.0093}_{-0.0056}$ & $0.7430^{+0.0126}_{-0.0070}$ & $0.742^{+0.039}_{-0.038}$ & This work \\
Density $\rho_*$ & $\rho_\odot$ & $2.397^{+0.017}_{-0.018}$ & $2.397^{+0.018}_{-0.019}$ & $2.40 \pm 0.31$ & This work \\
Luminosity $L_*$ & L$_\odot$ & $0.1682^{+0.0043}_{-0.0035}$ & $0.1664^{+0.0038}_{-0.0035}$ & $0.1673^{+0.0053}_{-0.0049}$ & This work \\
Projected rot. velocity $v\sin i$ & km s$^{-1}$ & $<2$ & - & - & This work \\
\tableline
\end{tabular}
\end{center}
\end{table*}

\section{Stellar Characterization} \label{stellar_char}

In order to characterize the star, we started by combining all of our collected HARPS-N spectra (from 2018-2019) into a single, stacked spectrum with S/N $\sim$ 300 (based on signal divided by scatter on continuum segments near 6000 \AA; see Section 3.1 of \citealt{mortieretal2013} for more details). Then we ran the ARESv2 package \citep{sousaetal2015} to obtain equivalent widths for a standard set of neutral and ionised iron lines \citep{sousaetal2011}. We refer to \citet{mortieretal2013}, \citet{sousa2014}, and \citet{sousaetal2015} for our choice of typical model parameters. Afterward, we calculated stellar parameters using MOOG\footnote[1]{2017 version: \url{http://www.as.utexas.edu/$\sim$chris/moog.html}} \citep{sneden1973} with ATLAS plane-parallel model atmospheres \citep{kurucz1993} assuming local thermodynamic equilibrium. A downhill simplex minimization procedure \citep{press1992} was used to determine the stellar photospheric parameters \citep[see e.g.][and references therein]{mortieretal2013}. We determined that the stellar temperature was less than $5200$ K, so we reran the minimization procedure with a sublist of lines designed for cooler stars \citep{tsantakietal2013}; we also constrained our line list to those with equivalent widths between $5$ and $150$ milliAngstroms (m\AA), removing $5$ lines above $150$ m\AA\ and $1$ line below $5$ m\AA\ (lines within this range tend to be sufficiently strong and well-described by a Gaussian). Finally, we corrected for $\log{g}$ and re-scaled errors following \citet{torresetal2012}, \citet{mortieretal2014}, and \citet{sousaetal2011}. The resulting effective temperature, surface gravity, microturbulence, and metallicity are reported in Table~\ref{stellar_table}.

Then we determined the same stellar parameters from the same spectra with a different, independent tool: the Stellar Parameter Classification tool (SPC; \citealt{buchhaveetal2012}). SPC interpolates across a synthetic spectrum library from \citet{kurucz1992} to find the best fit and uncertainties on an input spectrum. In addition to the stellar parameters calculated from ARES+MOOG, this tool also estimated rotational velocity. All atmospheric stellar parameters from ARES+MOOG and SPC were in good agreement (within $1\sigma$). Like ARES+MOOG, all SPC parameter estimates can be found in Table~\ref{stellar_table}.

We then took our estimated effective temperature and metallicity from ARES+MOOG and SPC, the \textit{Gaia} Data Release 3 (DR3) parallax \citep{gaiacollaboration2016b, gaiacollaboration2020, gaiacollaboration2022}, and numerous photometric magnitudes (B, V, J, H, K, W1, W2, and W3) and input them into the \texttt{isochrones} Python package \citep{morton2015a}. This package used two different sets of isochrones: Dartmouth \citep{dotteretal2008} and Modules for Experiments in Stellar Astrophysics (MESA) Isochrones and Stellar Tracks (MIST; \citealt{choietal2016,dotteretal2016}). Comparing two standard, independent models is useful for mitigating systematic errors and revealing discrepancies or issues in the resulting parameter estimates. We used \texttt{MultiNest} \citep{ferozetal2009,ferozetal2013} for parameter estimation, assuming 600 live points and otherwise standard \texttt{MultiNest} settings: importance nested sampling mode, multimodal mode, constant efficiency mode disabled, evidence tolerance $= 0.5$, and sampling efficiency $= 0.8$. As stated earlier, K2-136 is a member of the Hyades and therefore has a very tight age constraint of $650 \pm 70$ Myr \citep{martinetal2018}. We applied a much broader age prior of 475 Myr - 775 Myr (a $3\sigma$ range on the $625 \pm 50$ Hyades age estimate from \citealt{perrymanetal1998}), which was more than sufficient to achieve convergence. This yielded posterior distributions from both input atmospheric parameter sets (ARES+MOOG and SPC) as well as both isochrone sets (Dartmouth and MIST), for a total of four sets of posterior distributions (based on all combinations of input parameters and isochrones). 

The posteriors were then combined together (i.e. the posterior samples were appended together) to yield a single posterior distribution for each parameter. Lastly, systematic uncertainties determined by \citet{tayaretal2020} were added in quadrature to the combined posteriors to yield final parameters and uncertainties. Specifically, we added $4\%$ uncertainty to $R_\odot$, $5\%$ uncertainty to $M_\odot$, $2\%$ uncertainty to $L_\odot$, and $13\%$ uncertainty to $\rho_\odot$ (propagated from $R_\odot$ and $M_\odot$ uncertainties). The input \textit{Gaia} DR3 parallax, distance, photometric magnitudes, and the resulting stellar radius, mass, density, and luminosity are all reported in Table~\ref{stellar_table}.



\LongTables
\begin{longtable*}{cccccccccc}
\caption{\\RV observations and activity indicators\label{table:rv_table}}\\
\hline
\hline
BJD (TDB) & RV & $\sigma_{\mathrm{RV}}$ & CCF BIS & $\sigma_{\mathrm{BIS}}$ & CCF FWHM & $\sigma_{\mathrm{FWHM}}$ & $S_{HK}$ & $\sigma_{\mathrm{S_{HK}}}$ & Instrument \\
& (m s$^{-1}$) & (m s$^{-1}$) & (m s$^{-1}$) & (m s$^{-1}$) & (m s$^{-1}$) & (m s$^{-1}$) & & & \\
\hline\\
\endfirsthead

\caption{{\bf(continued)}\\RV observations and activity indicators\label{table:rv_table}}\\
\hline
\hline
BJD (TDB) & RV & $\sigma_{\mathrm{RV}}$ & CCF BIS & $\sigma_{\mathrm{BIS}}$ & CCF FWHM & $\sigma_{\mathrm{FWHM}}$ & $S_{HK}$ & $\sigma_{\mathrm{S_{HK}}}$ & Instrument \\
& (m s$^{-1}$) & (m s$^{-1}$) & (m s$^{-1}$) & (m s$^{-1}$) & (m s$^{-1}$) & (m s$^{-1}$) & & & \\
\hline\\
\endhead

\hline
\endfoot

$2458341.70618026$ & $39498.9$ & $1.8$ & $6901.1$ & $3.7$ & $63.8$ & $3.7$ & $1.191$ & $0.022$ & HARPS-N \\
$2458345.70159539$ & $39516.8$ & $1.4$ & $6920.6$ & $2.7$ & $63.4$ & $2.7$ & $1.198$ & $0.014$ & HARPS-N \\
$2458346.70175761$ & $39515.5$ & $2.0$ & $6948.0$ & $4.0$ & $70.2$ & $4.0$ & $1.274$ & $0.025$ & HARPS-N \\
$2458361.71645940$ & $39513.6$ & $2.9$ & $6945.1$ & $5.7$ & $75.3$ & $5.7$ & $1.234$ & $0.039$ & HARPS-N \\
$2458363.75019615$ & $39514.0$ & $1.4$ & $6942.8$ & $2.8$ & $69.2$ & $2.8$ & $1.222$ & $0.013$ & HARPS-N \\
$2458364.69673319$ & $39516.3$ & $1.4$ & $6941.9$ & $2.8$ & $74.6$ & $2.8$ & $1.251$ & $0.014$ & HARPS-N \\
$2458365.69775405$ & $39515.1$ & $1.3$ & $6949.1$ & $2.7$ & $71.9$ & $2.7$ & $1.238$ & $0.013$ & HARPS-N \\
$2458366.72988568$ & $39504.4$ & $1.7$ & $6932.3$ & $3.3$ & $83.8$ & $3.3$ & $1.212$ & $0.018$ & HARPS-N \\
$2458378.65078445$ & $39517.9$ & $2.2$ & $6947.9$ & $4.4$ & $72.8$ & $4.4$ & $1.294$ & $0.027$ & HARPS-N \\
$2458379.66372361$ & $39514.4$ & $1.4$ & $6945.1$ & $2.8$ & $80.4$ & $2.8$ & $1.249$ & $0.014$ & HARPS-N \\
$2458380.65930503$ & $39510.6$ & $1.2$ & $6930.6$ & $2.4$ & $77.3$ & $2.4$ & $1.294$ & $0.012$ & HARPS-N \\
$2458381.66442355$ & $39504.7$ & $1.2$ & $6921.7$ & $2.3$ & $86.1$ & $2.3$ & $1.251$ & $0.011$ & HARPS-N \\
$2458382.66425283$ & $39501.7$ & $1.6$ & $6913.0$ & $3.2$ & $78.8$ & $3.2$ & $1.243$ & $0.018$ & HARPS-N \\
$2458383.66210025$ & $39495.8$ & $2.3$ & $6902.9$ & $4.5$ & $75.2$ & $4.5$ & $1.172$ & $0.029$ & HARPS-N \\
$2458384.71464839$ & $39498.4$ & $2.2$ & $6896.4$ & $4.4$ & $67.9$ & $4.4$ & $1.160$ & $0.027$ & HARPS-N \\
$2458385.66982781$ & $39500.0$ & $1.4$ & $6900.7$ & $2.8$ & $63.5$ & $2.8$ & $1.162$ & $0.014$ & HARPS-N \\
$2458386.73771712$ & $39499.1$ & $1.5$ & $6910.7$ & $3.1$ & $59.3$ & $3.1$ & $1.165$ & $0.016$ & HARPS-N \\
$2458388.70001017$ & $39504.0$ & $1.4$ & $6916.2$ & $2.9$ & $62.5$ & $2.9$ & $1.213$ & $0.015$ & HARPS-N \\
$2458390.75891115$ & $39507.6$ & $1.5$ & $6923.3$ & $3.0$ & $66.3$ & $3.0$ & $1.194$ & $0.016$ & HARPS-N \\
$2458391.74334211$ & $39509.3$ & $1.5$ & $6907.4$ & $2.9$ & $65.3$ & $2.9$ & $1.203$ & $0.015$ & HARPS-N \\
$2458410.64445231$ & $39506.1$ & $2.2$ & $6903.3$ & $4.3$ & $82.2$ & $4.3$ & $1.165$ & $0.027$ & HARPS-N \\
$2458410.73761710$ & $39506.9$ & $2.7$ & $6904.2$ & $5.4$ & $81.5$ & $5.4$ & $1.143$ & $0.037$ & HARPS-N \\
$2458415.61383639$ & $39507.7$ & $2.4$ & $6905.6$ & $4.8$ & $66.5$ & $4.8$ & $1.130$ & $0.028$ & HARPS-N \\
$2458415.72038413$ & $39504.5$ & $1.4$ & $6916.4$ & $2.9$ & $68.7$ & $2.9$ & $1.163$ & $0.015$ & HARPS-N \\
$2458421.64508561$ & $39500.0$ & $1.4$ & $6937.6$ & $2.8$ & $74.8$ & $2.8$ & $1.240$ & $0.014$ & HARPS-N \\
$2458421.72295016$ & $39497.8$ & $4.0$ & $6953.8$ & $8.0$ & $67.1$ & $8.0$ & $1.317$ & $0.062$ & HARPS-N \\
$2458424.69841781$ & $39505.2$ & $4.4$ & $6922.8$ & $8.8$ & $63.0$ & $8.8$ & $1.255$ & $0.068$ & HARPS-N \\
$2458424.76629848$ & $39502.2$ & $2.9$ & $6916.1$ & $5.7$ & $69.5$ & $5.7$ & $1.247$ & $0.041$ & HARPS-N \\
$2458448.58055249$ & $39506.0$ & $1.2$ & $6911.5$ & $2.3$ & $63.8$ & $2.3$ & $1.137$ & $0.010$ & HARPS-N \\
$2458448.71120394$ & $39506.1$ & $1.8$ & $6917.5$ & $3.6$ & $75.5$ & $3.6$ & $1.160$ & $0.020$ & HARPS-N \\
$2458449.42303120$ & $39518.1$ & $5.9$ & $6925$ & $12$ & $81$ & $12$ & $1.17$ & $0.11$ & HARPS-N \\
$2458449.69543641$ & $39507.7$ & $2.3$ & $6903.0$ & $4.5$ & $63.2$ & $4.5$ & $1.158$ & $0.029$ & HARPS-N \\
$2458451.47509484$ & $39511.7$ & $1.6$ & $6923.5$ & $3.2$ & $69.2$ & $3.2$ & $1.164$ & $0.017$ & HARPS-N \\
$2458451.61304290$ & $39512.1$ & $1.5$ & $6917.5$ & $2.9$ & $68.5$ & $2.9$ & $1.184$ & $0.015$ & HARPS-N \\
$2458453.60424787$ & $39495.9$ & $1.6$ & $6922.3$ & $3.3$ & $79.5$ & $3.3$ & $1.242$ & $0.019$ & HARPS-N \\
$2458453.70820393$ & $39493.9$ & $1.6$ & $6914.4$ & $3.3$ & $78.0$ & $3.3$ & $1.173$ & $0.019$ & HARPS-N \\
$2458454.46507987$ & $39492.6$ & $1.5$ & $6902.0$ & $3.1$ & $75.7$ & $3.1$ & $1.155$ & $0.017$ & HARPS-N \\
$2458454.55426377$ & $39488.9$ & $1.5$ & $6907.1$ & $3.0$ & $77.3$ & $3.0$ & $1.180$ & $0.015$ & HARPS-N \\
$2458456.47471787$ & $39499.7$ & $1.5$ & $6895.0$ & $3.0$ & $68.5$ & $3.0$ & $1.198$ & $0.016$ & HARPS-N \\
$2458462.64638556$ & $39509.7$ & $1.7$ & $6915.3$ & $3.4$ & $60.7$ & $3.4$ & $1.237$ & $0.021$ & HARPS-N \\
$2458473.54081960$ & $39494.9$ & $1.3$ & $6911.9$ & $2.5$ & $73.2$ & $2.5$ & $1.092$ & $0.011$ & HARPS-N \\
$2458473.63947550$ & $39492.6$ & $1.3$ & $6909.0$ & $2.7$ & $74.8$ & $2.7$ & $1.068$ & $0.012$ & HARPS-N \\
$2458474.47139808$ & $39494.1$ & $1.3$ & $6913.1$ & $2.7$ & $76.0$ & $2.7$ & $1.067$ & $0.012$ & HARPS-N \\
$2458474.56083109$ & $39491.6$ & $1.1$ & $6911.6$ & $2.2$ & $66.0$ & $2.2$ & $1.1493$ & $0.0099$ & HARPS-N \\
$2458477.50294855$ & $39506.5$ & $2.9$ & $6915.2$ & $5.8$ & $70.1$ & $5.8$ & $1.164$ & $0.039$ & HARPS-N \\
$2458477.60820837$ & $39503.8$ & $3.4$ & $6950.6$ & $6.7$ & $64.1$ & $6.7$ & $1.120$ & $0.048$ & HARPS-N \\
$2458478.43374093$ & $39504.7$ & $1.9$ & $6914.6$ & $3.8$ & $66.9$ & $3.8$ & $1.164$ & $0.021$ & HARPS-N \\
$2458478.55045568$ & $39502.0$ & $1.3$ & $6917.9$ & $2.7$ & $72.5$ & $2.7$ & $1.250$ & $0.013$ & HARPS-N \\
$2458479.54013997$ & $39497.3$ & $1.1$ & $6919.3$ & $2.2$ & $65.9$ & $2.2$ & $1.223$ & $0.010$ & HARPS-N \\
$2458479.59583870$ & $39498.3$ & $1.2$ & $6908.3$ & $2.4$ & $69.5$ & $2.4$ & $1.235$ & $0.012$ & HARPS-N \\
$2458480.49811363$ & $39498.5$ & $1.6$ & $6904.6$ & $3.3$ & $80.0$ & $3.3$ & $1.164$ & $0.017$ & HARPS-N \\
$2458480.61833460$ & $39500.4$ & $3.3$ & $6922.6$ & $6.5$ & $74.6$ & $6.5$ & $1.182$ & $0.049$ & HARPS-N \\
$2458481.52056002$ & $39494.5$ & $1.9$ & $6896.6$ & $3.8$ & $74.5$ & $3.8$ & $1.150$ & $0.022$ & HARPS-N \\
$2458481.61815376$ & $39500.0$ & $1.5$ & $6893.8$ & $2.9$ & $74.9$ & $2.9$ & $1.174$ & $0.016$ & HARPS-N \\
$2458482.47714376$ & $39502.2$ & $1.4$ & $6893.7$ & $2.8$ & $68.9$ & $2.8$ & $1.127$ & $0.014$ & HARPS-N \\
$2458482.57623761$ & $39502.2$ & $1.2$ & $6888.0$ & $2.4$ & $67.9$ & $2.4$ & $1.135$ & $0.011$ & HARPS-N \\
$2458483.48217046$ & $39506.5$ & $1.4$ & $6892.5$ & $2.8$ & $66.2$ & $2.8$ & $1.120$ & $0.013$ & HARPS-N \\
$2458483.57932245$ & $39507.8$ & $1.3$ & $6886.6$ & $2.6$ & $70.6$ & $2.6$ & $1.118$ & $0.012$ & HARPS-N \\
$2458484.45739679$ & $39506.9$ & $1.8$ & $6894.0$ & $3.6$ & $61.6$ & $3.6$ & $1.160$ & $0.020$ & HARPS-N \\
$2458484.56221664$ & $39507.2$ & $1.5$ & $6892.7$ & $2.9$ & $71.3$ & $2.9$ & $1.127$ & $0.015$ & HARPS-N \\
$2458486.56813502$ & $39505.3$ & $2.0$ & $6905.7$ & $4.1$ & $75.7$ & $4.1$ & $1.221$ & $0.027$ & HARPS-N \\
$2458487.44451438$ & $39500.6$ & $1.4$ & $6899.3$ & $2.8$ & $69.1$ & $2.8$ & $1.192$ & $0.015$ & HARPS-N \\
$2458487.55454800$ & $39501.7$ & $1.4$ & $6901.2$ & $2.8$ & $69.6$ & $2.8$ & $1.170$ & $0.015$ & HARPS-N \\
$2458488.41884010$ & $39497.3$ & $1.2$ & $6901.9$ & $2.5$ & $66.8$ & $2.5$ & $1.138$ & $0.012$ & HARPS-N \\
$2458488.52873713$ & $39499.6$ & $1.9$ & $6900.0$ & $3.9$ & $66.3$ & $3.9$ & $1.206$ & $0.024$ & HARPS-N \\
$2458489.43534717$ & $39496.3$ & $1.8$ & $6899.7$ & $3.6$ & $70.6$ & $3.6$ & $1.193$ & $0.021$ & HARPS-N \\
$2458489.58136198$ & $39492.8$ & $1.6$ & $6908.5$ & $3.1$ & $67.7$ & $3.1$ & $1.166$ & $0.018$ & HARPS-N \\
$2458502.38887026$ & $39505.0$ & $1.7$ & $6903.9$ & $3.3$ & $66.2$ & $3.3$ & $1.123$ & $0.018$ & HARPS-N \\
$2458502.53993777$ & $39505.4$ & $2.5$ & $6896.3$ & $5.0$ & $77.0$ & $5.0$ & $1.120$ & $0.032$ & HARPS-N \\
$2458503.37914022$ & $39506.8$ & $1.5$ & $6904.6$ & $2.9$ & $71.5$ & $2.9$ & $1.183$ & $0.015$ & HARPS-N \\
$2458503.50487143$ & $39505.3$ & $1.2$ & $6903.0$ & $2.5$ & $74.4$ & $2.5$ & $1.170$ & $0.012$ & HARPS-N \\
$2458504.43989909$ & $39509.9$ & $2.4$ & $6912.6$ & $4.8$ & $81.1$ & $4.8$ & $1.114$ & $0.029$ & HARPS-N \\
$2458504.55233098$ & $39503.5$ & $2.1$ & $6915.0$ & $4.2$ & $70.0$ & $4.2$ & $1.102$ & $0.025$ & HARPS-N \\
$2458505.38129467$ & $39500.6$ & $1.7$ & $6905.6$ & $3.3$ & $72.6$ & $3.3$ & $1.200$ & $0.019$ & HARPS-N \\
$2458505.49840229$ & $39498.6$ & $2.3$ & $6909.7$ & $4.7$ & $77.7$ & $4.7$ & $1.118$ & $0.030$ & HARPS-N \\
$2458506.36962284$ & $39493.9$ & $3.6$ & $6916.2$ & $7.1$ & $66.2$ & $7.1$ & $1.243$ & $0.055$ & HARPS-N \\
$2458506.50672836$ & $39488.8$ & $2.9$ & $6899.0$ & $5.8$ & $71.8$ & $5.8$ & $1.171$ & $0.041$ & HARPS-N \\
$2458518.37653154$ & $39508.5$ & $2.2$ & $6909.8$ & $4.3$ & $66.3$ & $4.3$ & $1.193$ & $0.027$ & HARPS-N \\
$2458518.45830362$ & $39515.6$ & $4.3$ & $6932.0$ & $8.7$ & $73.9$ & $8.7$ & $1.278$ & $0.072$ & HARPS-N \\
$2458518.47977117$ & $39509.4$ & $2.8$ & $6914.6$ & $5.5$ & $76.9$ & $5.5$ & $1.203$ & $0.040$ & HARPS-N \\
$2458519.35187655$ & $39498.8$ & $3.8$ & $6910.8$ & $7.7$ & $74.4$ & $7.7$ & $1.120$ & $0.061$ & HARPS-N \\
$2458519.45005233$ & $39502.4$ & $3.5$ & $6921.3$ & $7.0$ & $84.8$ & $7.0$ & $1.213$ & $0.055$ & HARPS-N \\
$2458520.35461693$ & $39497.1$ & $1.5$ & $6893.9$ & $2.9$ & $74.7$ & $2.9$ & $1.142$ & $0.015$ & HARPS-N \\
$2458520.45526091$ & $39494.5$ & $1.8$ & $6914.5$ & $3.5$ & $72.8$ & $3.5$ & $1.157$ & $0.021$ & HARPS-N \\
$2458521.40212719$ & $39499.0$ & $2.2$ & $6889.8$ & $4.5$ & $71.3$ & $4.5$ & $1.183$ & $0.029$ & HARPS-N \\
$2458521.48828747$ & $39498.2$ & $2.1$ & $6899.6$ & $4.2$ & $71.1$ & $4.2$ & $1.147$ & $0.027$ & HARPS-N \\
$2458522.35495354$ & $39499.9$ & $1.2$ & $6887.3$ & $2.4$ & $65.3$ & $2.4$ & $1.132$ & $0.011$ & HARPS-N \\
$2458522.43817942$ & $39500.6$ & $1.4$ & $6892.8$ & $2.8$ & $67.9$ & $2.8$ & $1.130$ & $0.015$ & HARPS-N \\
$2458788.78775712$ & $39499.17$ & $0.89$ & $6953.5$ & $1.8$ & $43.7$ & $1.8$ & $1.1403$ & $0.0020$ & ESPRESSO \\
$2458804.67601741$ & $39480.32$ & $0.67$ & $6965.9$ & $1.3$ & $65.6$ & $1.3$ & $1.2545$ & $0.0013$ & ESPRESSO \\
$2458806.80347928$ & $39484.42$ & $0.79$ & $6928.5$ & $1.6$ & $56.1$ & $1.6$ & $1.0794$ & $0.0016$ & ESPRESSO \\
$2458808.76852035$ & $39504.65$ & $0.83$ & $6941.7$ & $1.7$ & $41.2$ & $1.7$ & $1.0573$ & $0.0018$ & ESPRESSO \\
$2458820.76166121$ & $39489.22$ & $0.46$ & $6928.21$ & $0.91$ & $48.14$ & $0.91$ & $1.18743$ & $0.00067$ & ESPRESSO \\
$2458825.59389775$ & $39492.79$ & $0.71$ & $6962.1$ & $1.4$ & $60.4$ & $1.4$ & $1.1525$ & $0.0014$ & ESPRESSO \\
$2458833.58700359$ & $39486.65$ & $0.70$ & $6934.8$ & $1.4$ & $54.4$ & $1.4$ & $1.1807$ & $0.0013$ & ESPRESSO \\
$2458839.65413441$ & $39493.99$ & $0.63$ & $6955.9$ & $1.3$ & $54.1$ & $1.3$ & $1.2941$ & $0.0011$ & ESPRESSO \\
$2458840.58619720$ & $39501.76$ & $0.70$ & $6967.2$ & $1.4$ & $41.1$ & $1.4$ & $1.3193$ & $0.0013$ & ESPRESSO \\
$2458848.60085991$ & $39501.35$ & $0.49$ & $6942.72$ & $0.98$ & $43.38$ & $0.98$ & $1.17686$ & $0.00075$ & ESPRESSO \\
$2458849.56597102$ & $39505.53$ & $0.69$ & $6955.6$ & $1.4$ & $38.6$ & $1.4$ & $1.1913$ & $0.0013$ & ESPRESSO \\
$2458850.58225768$ & $39494.9$ & $1.8$ & $6972.4$ & $3.5$ & $46.6$ & $3.5$ & $1.2761$ & $0.0044$ & ESPRESSO \\
$2458850.60464655$ & $39497.81$ & $0.92$ & $6955.0$ & $1.8$ & $34.1$ & $1.8$ & $1.2108$ & $0.0019$ & ESPRESSO \\
$2458851.58865177$ & $39493.17$ & $0.79$ & $6953.3$ & $1.6$ & $57.3$ & $1.6$ & $1.2570$ & $0.0016$ & ESPRESSO \\
$2458853.68186920$ & $39502.19$ & $0.57$ & $6968.8$ & $1.1$ & $58.0$ & $1.1$ & $1.3340$ & $0.0010$ & ESPRESSO \\
$2458864.55947341$ & $39502.03$ & $0.65$ & $6954.6$ & $1.3$ & $44.6$ & $1.3$ & $1.2129$ & $0.0012$ & ESPRESSO \\
$2458864.64168610$ & $39502.17$ & $0.77$ & $6944.7$ & $1.5$ & $53.7$ & $1.5$ & $1.2462$ & $0.0017$ & ESPRESSO \\
$2458865.59784818$ & $39505.73$ & $0.94$ & $6978.3$ & $1.9$ & $38.7$ & $1.9$ & $1.3126$ & $0.0024$ & ESPRESSO \\
$2458869.61564757$ & $39488.28$ & $0.56$ & $6996.3$ & $1.1$ & $63.6$ & $1.1$ & $1.2687$ & $0.0011$ & ESPRESSO \\
$2458886.57544506$ & $39481.55$ & $0.69$ & $6924.4$ & $1.4$ & $52.8$ & $1.4$ & $1.2086$ & $0.0013$ & ESPRESSO \\
$2458887.57037120$ & $39487.19$ & $0.61$ & $6925.7$ & $1.2$ & $45.6$ & $1.2$ & $1.1117$ & $0.0011$ & ESPRESSO \\
$2458906.52291945$ & $39500.21$ & $0.67$ & $6988.0$ & $1.3$ & $45.9$ & $1.3$ & $1.2254$ & $0.0012$ & ESPRESSO \\
$2459110.65183858$ & $39492.2$ & $2.7$ & $6957.0$ & $5.3$ & $66.3$ & $5.3$ & $1.251$ & $0.033$ & HARPS-N \\
$2459111.66733900$ & $39505.2$ & $1.2$ & $6956.5$ & $2.5$ & $67.6$ & $2.5$ & $1.401$ & $0.011$ & HARPS-N \\
$2459112.66195511$ & $39508.2$ & $1.9$ & $6955.8$ & $3.8$ & $64.1$ & $3.8$ & $1.294$ & $0.020$ & HARPS-N \\
$2459120.71102749$ & $39512.7$ & $1.3$ & $6979.6$ & $2.7$ & $70.1$ & $2.7$ & $1.339$ & $0.013$ & HARPS-N \\
$2459153.51589766$ & $39502.7$ & $2.4$ & $6939.2$ & $4.7$ & $67.1$ & $4.7$ & $1.216$ & $0.026$ & HARPS-N \\
\end{longtable*}

\subsection{Stellar Rotation Period}\label{rotation_period}

One parameter of special interest is the stellar rotation period, which we include as a parameter in our RV model (see Section~\ref{rv_model}). M18 conducted a Lomb-Scargle periodogram on the \emph{K2} light curve and reported a rotation period of $15.04 \pm 1.01$ days. \citet{ciardietal2018} analyzed the same light curve and found a rotation period of $15.2 \pm 0.2$ days through a Lomb-Scargle periodogram and $13.8 \pm 1.0$ days through an autocorrelation function. \citet{livingstonetal2018} conducted a Gaussian process (GP) regression, a Lomb-Scargle periodogram, and an autocorrelation function on the light curve and found a corresponding rotation period of $13.5^{+0.7}_{-0.4}$ d, $15.1^{+1.3}_{-1.2}$ d, and $13.6^{+2.2}_{-1.5}$ d, respectively. Note: Given an offset and uncertainties in the photometric data set, a generalized Lomb-Scargle periodogram would be preferred \citep{zechmeisterandkurster2009}; however, it is not clear from the referenced papers whether this generalized method was used or just a basic Lomb-Scargle periodogram \citep{scargle1982}.

Notably, the estimates via a Lomb-Scargle periodogram are longer than estimates with other methods. All results are broadly consistent with our findings from our full model results ($13.37^{+0.13}_{-0.17}$ days; see Table~\ref{rv_results}), except the $15.2 \pm 0.2$ day result from the Lomb-Scargle analysis by \citet{ciardietal2018}. They also have the smallest uncertainties of any rotation period estimate, so it is possible that their value is reasonable but the uncertainties are overly optimistic. 

A possible explanation of this discrepancy could be differential rotation. Regardless of activity level, starspots, plage, and other activity may be more prominent at different stellar latitudes when the \emph{K2} photometry and our HARPS-N spectroscopy were conducted. This hypothesis is also mentioned by \citet{ciardietal2018} to explain a larger than expected v$\sin{i}$. Following \citet{barnesetal2005} and \citet{kitchatinovetal2012}, they estimate that the equatorial rotation period of K2-136 could be faster than higher latitudes by $\sim1$ day. Then again, \citet{aigrainetal2015} found that claims of differential rotation should be treated with caution even for long baselines of photometry. We may simply be seeing different starspots at different longitudes creating phase modulation, combined with greater or fewer numbers of starspots leading to better or worse constraints on rotation period.

\subsection{Binarity of K2-136}\label{binarity}

One of the planet discovery papers, \citet{ciardietal2018}, reported a binary companion to K2-136. In addition to their \emph{K2} photometric analysis, they collected spectra from the SpeX spectrograph \citep{rayneretal2003,rayneretal2004} at the 3-m NASA Infrared Telescope Facility and the HIRES spectrograph \citep{vogtetal1994} at the Keck I telescope, as well as AO observations with the NIRC2 instrument at the Keck II telescope and the P3K AO system and PHARO camera \citep{hawyardetal2001} on the 200'' Hale Telescope at Palomar Observatory. The AO observations at both facilities detected an M7/8V star separated from the primary star by $\sim0.7$'', corresponding to a projected separation of $\sim 40$ AU; the spectroscopic observations did not detect this companion, and no further companions were found by any of the above observations. (Notably, this angular separation is more than enough for HST to resolve; see Sections~\ref{HST_discussion} and ~\ref{HST_obs}.)

Gaia DR3 did not detect the binary companion, leaving the issue of boundedness unresolved. However, \citet{ciardietal2018} compared the current position of K2-136 against observations from the 1950 Palomar Observatory Sky Survey (POSS I) and noted that the star had moved $6$'' in the intervening time with no evidence of background stars. These POSS I observations show that the stellar companion is likely bound.

Further, Gaia DR3 reported K2-136 to have an astrometric excess noise of 96 $\mu$as and a Renormalised Unit Weight Error (RUWE) of 1.23, a mild departure from a good single-star model. At the separation and brightness of the companion, this excess variability in the astrometry is unlikely to be due to pollution from its light contribution: at $\sim0.7$'' separation a companion can be detected only with a G magnitude difference of $\lesssim 2$ \citep{gaiacollaboration2021}. There is therefore a mild indication of astrometric variability due to unmodeled orbital motion. Using the formalism of \citet{torresetal1999}, at the distance of the system, given its angular separation, and for the mass range of an M7/8 star, the median astrometric acceleration is expected to be $\sim25$ $\mu$as yr$^{-2}$, with maximum value close to $40$ $\mu$as yr$^{-2}$, indicating that in addition to simple astrometric noise, the bulk of the astrometric variability could be caused by the detection of the acceleration due to the companion.

It is worth considering whether flux from the companion could bias the measured RVs of the primary K dwarf. Both the HARPS-N and ESPRESSO band passes are approximately $380$nm – $690$nm and centered on the V band. According to \citet{ciardietal2018}, the M dwarf companion is at least $10$ magnitudes fainter than the primary in the V-band. We can use $\Delta{m} = 10$ as a worst-case scenario and similarly assume the companion star was well-centered on the fiber for all observations (because the companion and primary are separated by $0.7$'', the companion would not be well-centered and the actual flux contamination from the companion would be less). \citet{cunhaetal2013} explored the RV impact of flux contamination from a stellar companion: for a K5 dwarf and an M dwarf (M3 or later) with $\Delta{m} = 10$, the maximum impact on RVs is $<10$ cm s$^{-1}$, and therefore negligible for our level of RV precision.

\citet{ciardietal2018} explored whether the transit signals may originate from the M dwarf. They found that in order to match the observed transit depth of K2-136c, the M dwarf would have to be a binary system itself that exhibits significant and detectable secondary eclipses, which have not been observed. Further, the transit duration of K2-136c is inconsistent with a transit of an M dwarf. Finally, K2-136c has already been validated by \citet{ciardietal2018} and all three planets have been independently validated by M18 and \citet{livingstonetal2018}. Therefore, it is very unlikely that the planets are false positive signals or planetary signals from the M dwarf.

However, in the \emph{Kepler} band pass, the companion M dwarf is 6.5 magnitudes fainter, which leads to a very small dilution effect on the planet transit depths. Following \citet{ciardietal2015}, we include this effect for the sake of robustness in our final planet radius estimates (which enlarges each planet by $\sim0.13\%$).

This binary companion may also cause a long-term RV trend, which we discuss further in Section~\ref{rv_model} and test in Section~\ref{trend_analysis}.

\begin{figure*}[ht!]
\epsscale{1.1}
  \begin{center}
      \leavevmode
\plotone{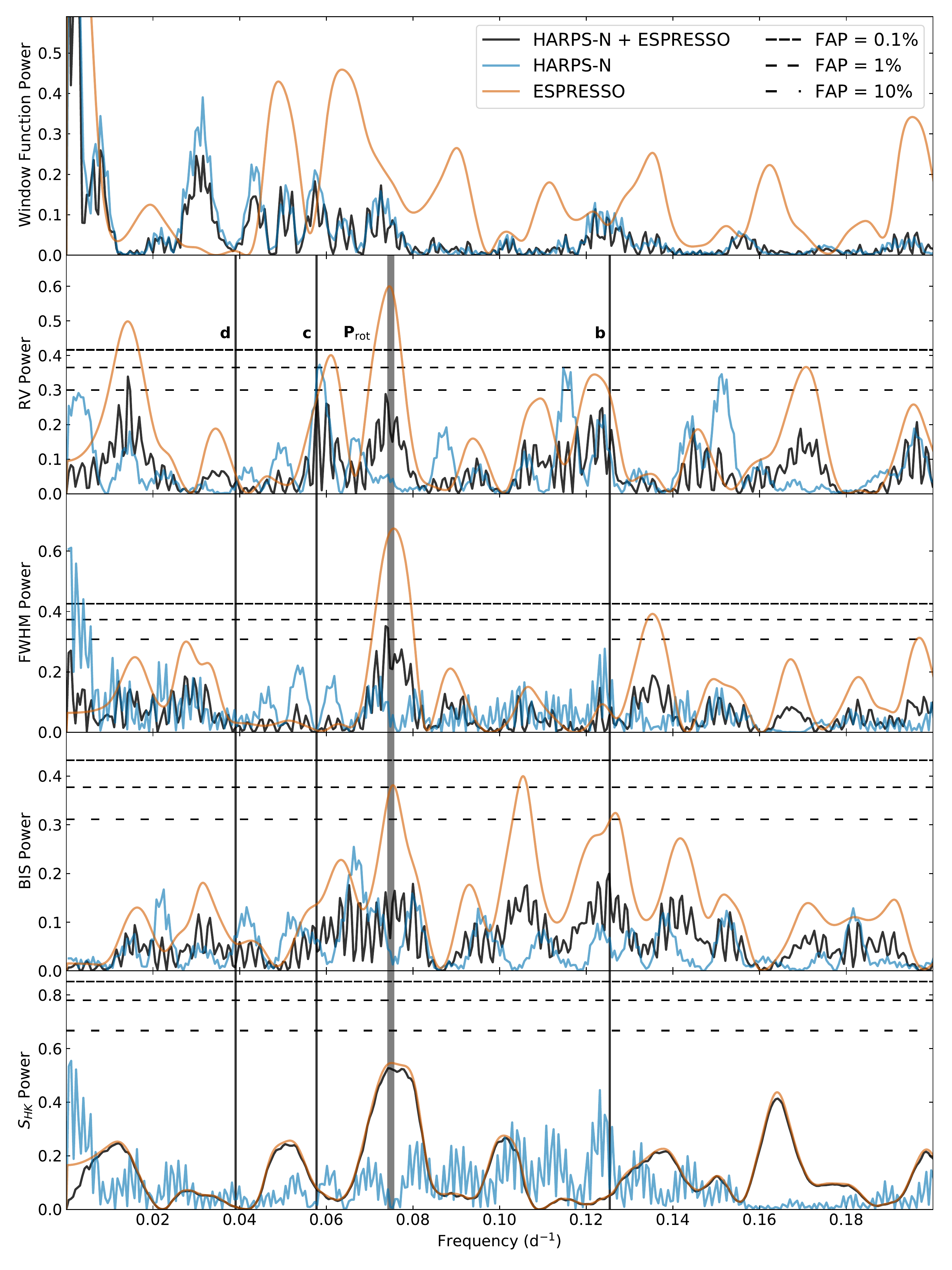}
\caption{Periodograms of RV, CCF FWHM, CCF BIS, and $S_{HK}$ for the K2-136 system. In the top panel is the window function (computed from observation times only). Each subplot has the periodogram of HARPS-N and ESPRESSO combined (black), HARPS-N alone (blue), and ESPRESSO alone (orange). The gray region corresponds to the $1\sigma$ confidence interval of the stellar rotation period (as determined from our model results); the three vertical, black lines correspond to the orbital periods of K2-136b, c, and d. Finally, the horizontal dashed lines refer to different false alarm probabilities (for HARPS-N and ESPRESSO combined).} \label{periodogram}
\end{center}
\end{figure*}

\section{Data Analysis} \label{analysis}

We analyzed the RV data in conjunction with a number of other common stellar activity indices that are calculated with the ESPRESSO DRS 2.2.8 pipeline, specifically the CCF BIS, CCF FWHM, and $S_{HK}$. In this section, we first explore the data set by generating a periodogram and conducting a correlation analysis between the RVs and other data types. Then we discuss the transit and RV components of our model, the parameter estimation process, and how we compare our models. Finally, we conduct tests on our results and discuss the implications of a binary companion in the system.

\begin{figure}[ht!]
\epsscale{1.1}
  \begin{center}
      \leavevmode
\plotone{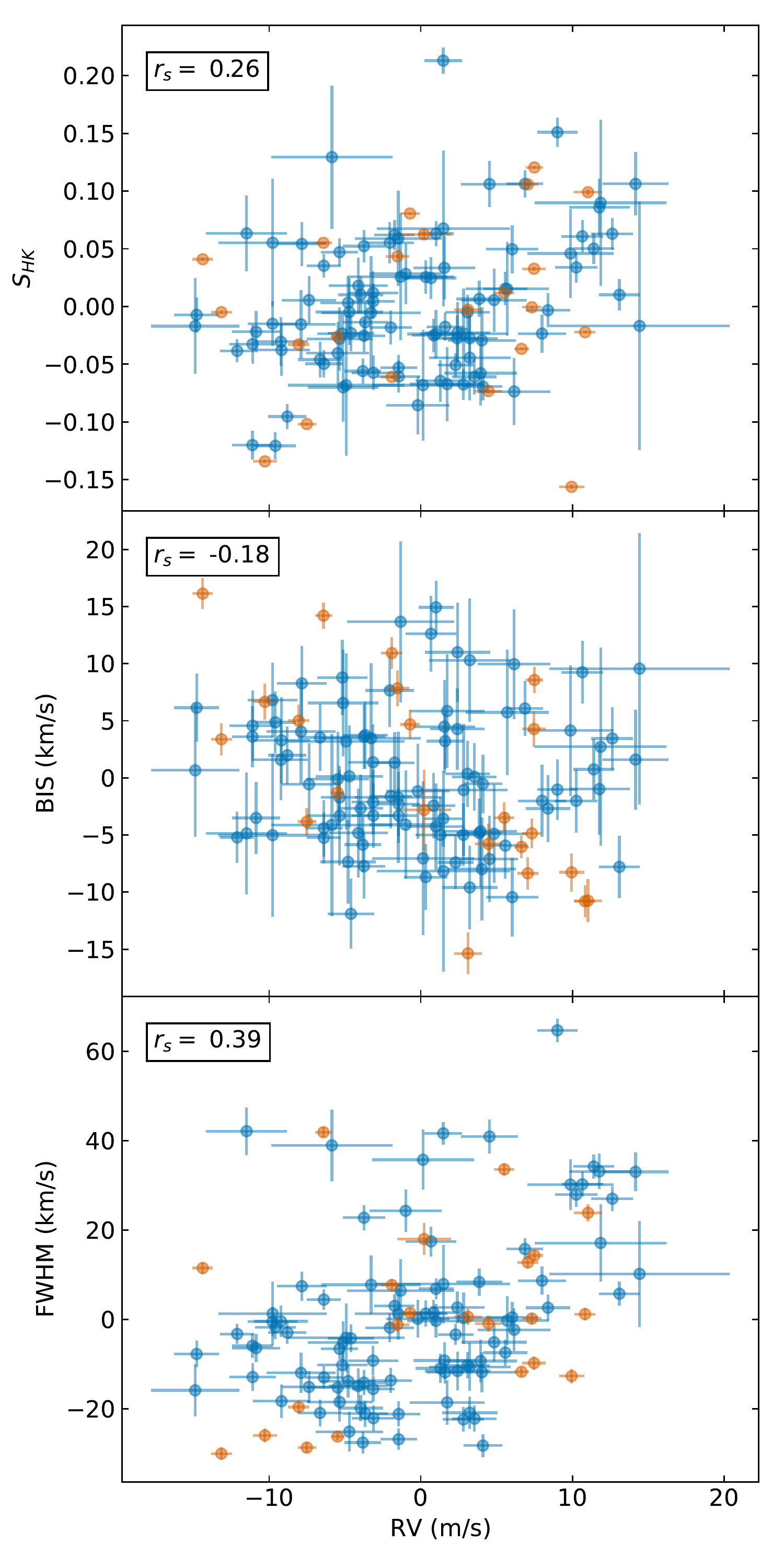}
\caption{Scatter plots of $S_{HK}$, BIS, and FWHM against RV for the K2-136 system. All HARPS-N and ESPRESSO data have been separately offset shifted according to the median model offsets listed in Table~\ref{rv_results}. Blue data points correspond to HARPS-N observations, orange data points to ESPRESSO. In the top-left corner of each subplot is the Spearman correlation coefficient, which can capture nonlinear, monotonic correlations. (Coefficients were calculated with data values but not uncertainties.)} \label{correlations}
\end{center}
\end{figure}

\subsection{Periodogram Analysis}
In order to investigate periodic signals in our data (planetary or otherwise), we created Generalized Lomb-Scargle periodograms \citep{scargle1982, zechmeisterandkurster2009} of our RVs and our stellar activity indices. As a point of reference, we also included the window function of our data (built from constant, non-zero values at each of the timestamps of our observations). It is used to determine the regular patterns in the periodograms due to the sampling and gaps in the time series. These periodograms are presented in Fig.~\ref{periodogram}.

To ascertain the robustness of any apparent signals, we also estimated each periodogram's False Alarm Probability (FAP), the likelihood that an apparent signal of a given strength will be detected when no underlying signal is actually present. The FAP was estimated with the bootstrap method: sampling the observations randomly with replacement while maintaining the same timestamps. We repeat this process 100000 times, each time constructing a periodogram and determining the maximum peak. This reveals how often a given signal strength will appear due only to noise, from which the FAP is calculated.

The strongest RV signals in our combined periodogram (both HARPS-N and ESPRESSO) are at the orbital period of K2-136c, the rotation period of the star, and near $0.017$ d$^{-1}$ (i.e. half the length of the ESPRESSO data baseline of $117.7$ days), although none are significant at the $1\%$ level. Among the stellar activity periodograms, the strongest signals are the rotation period signal in the ESPRESSO FWHM power and a long-period signal in the HARPS-N FWHM power which can be attributed to the window function. Also, the strongest peak in the (combined data) window function periodogram above $0.01$ days$^{-1}$ (near $0.03$ days$^{-1}$) does not correspond to any significant signals or aliases in any of the four data types. As for low-frequency signals $< 0.01$ days$^{-1}$, we discuss possible long-term trends in Sections~\ref{binarity} and \ref{rv_model}. All of this indicates that K2-136c has a more detectable RV signal than the other two planets, as expected given its size.

\subsection{Correlation Analysis}
We also examined the relationship between RVs and our stellar activity indices. Because RVs are measured from small shifts in spectral absorption lines, and stellar activity can change the shape of absorption lines, stellar activity can significantly affect RV observations \citep[e.g.][]{quelozetal2001, haywood2015, rajpauletal2015}. Scatter plots between RVs and all three activity indices are presented in Fig.~\ref{correlations}. At least in the case of $S_{HK}$ and FWHM, there are notable correlations with RVs according to the p-values for the Spearman correlation coefficient (which captures nonlinear, monotonic correlations and uses the same -1 to 1 range as the Pearson correlation coefficient). According to the Spearman coefficient, there is a correlation of $0.26$ between RV and $S_{HK}$, a negative correlation of $-0.18$ between RV and BIS, and the strongest correlation of $0.39$ between RV and FWHM. For a data set of this size, these coefficients correspond to p-values of $0.004$ for RV and $S_{HK}$, $0.052$ for RV and BIS, and $\ll 0.001$ for RV and FWHM. Therefore, for RV and $S_{HK}$ and especially RV and FWHM, there appears to be a statistically significant correlation. In other words, there is good reason to believe that this data set includes correlated and structured stellar activity. In fact, the Spearman correlation coefficient is likely an underestimate of the correlation between RVs and stellar activity indices, since there can be a phase shift in the amplitude variation from one data type to another \citep{santosetal2014,collier-cameronetal2019}.

\subsection{K2 Transit Photometry}\label{transit_model}

We cleaned and flattened the photometric data from \emph{K2} using the exact same procedure as was used originally in M18. Their procedure follows the self-flat-fielding (SFF) method developed in \citet{vanderburgandjohnson2014} to perform a rough removal of instrumental variability followed by a simultaneous fit to a model consisting of \citet{mandelandagol2002} transit shapes for the three planets, a basis spline in time to describe the stellar variability, and splines in \Kepler's roll angle to describe the systematic photometric errors introduced by the spacecraft's unstable pointing \citep{vanderburgetal2016a}. After performing the fit, they removed the best-fit systematics and stellar variability components, isolating the transits for further analysis. The interested reader should refer to M18 for additional detail of the full procedure.

\begin{figure*}[ht!]
\epsscale{1.17}
  \begin{center}
      \leavevmode
\plotone{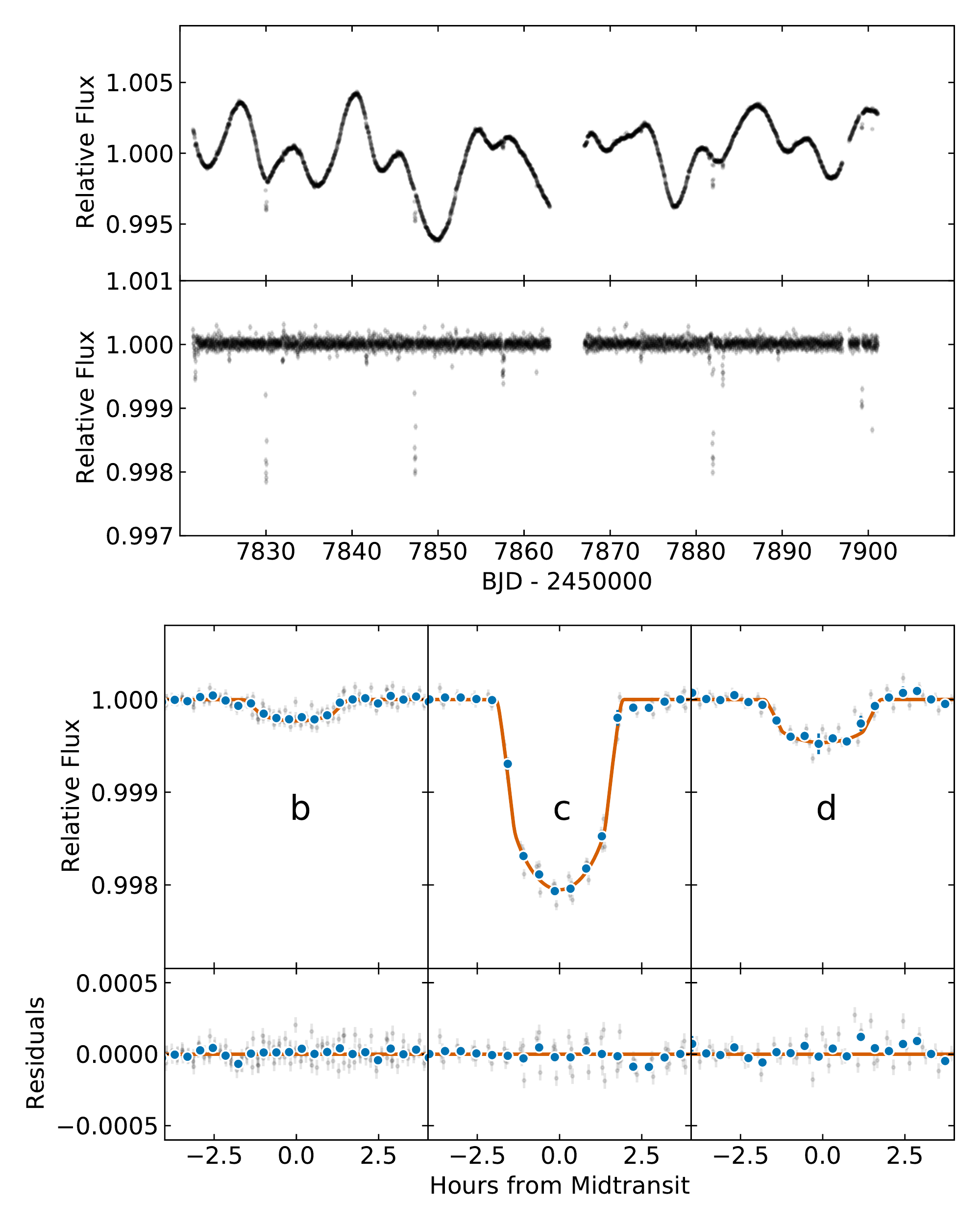}
\caption{Transit plot of K2-136. The top and bottom subplots of the top plot are the raw and normalized \emph{K2} photometry versus time from Campaign 13, respectively. In the top subplots of the bottom plot are the phase-folded light curves and transit model fits for K2-136b, c, and d. The gray points are the raw data. The best-fit transit model is the orange line and binning is represented by the blue points. The bottom subplots of the bottom plot are the residuals after the best-fit model has been subtracted.} \label{transits}
\end{center}
\end{figure*}

We modeled the flattened and cleaned M18 light curve with the \texttt{BATMAN} Python package \citep{kreidberg2015}, based on the \citet{mandelandagol2002} transit model. Our model included a baseline offset parameter and white noise parameter for our \emph{K2} Campaign 13 photometry as well as two quadratic limb-darkening parameters (parameterized using \citealt{kipping2013b}). Each planet was modeled with five parameters: the transit time, orbital period, planet radius relative to stellar radius, transit duration, and impact parameter. All parameters were modeled with either uniform, Gaussian, Jeffreys, or modified Jeffreys priors \citep{gregory2007}. Only the photometric white noise parameters used modified Jeffreys priors, with a knee located at the mean of the photometric flux uncertainty for that particular campaign or sector. All priors are listed in Table~\ref{transit_results}. The raw and flattened data can be seen in Fig.~\ref{transits}.

We also applied a Gaussian prior on stellar density by comparing the spectroscopically derived stellar density to the stellar density found via the following equation \citep{seagerandmallen-ornelas2003, sozzettietal2007}:

\begin{equation} \label{eqn:density}
    \rho_* = \frac{3\pi}{GP^2}\bigg(\frac{a}{R_*}\bigg)^3
\end{equation}

where orbital period ($P$) and the semi-major axis ($a/R_*$) are derived directly from the light curve model.

All together, our full transit model includes 15 planetary parameters (5 per planet: time of transit, orbital period, ratio of planet radius to stellar radius, transit duration, and impact parameter), 2 quadratic limb darkening parameters, 1 photometric noise parameter, and 1 photometric baseline offset parameter for a total of 19 parameters.

The only parameters in common between our transit model and our RV model (as explained in further detail below) are transit times and orbital periods.

\subsection{TESS Transit Photometry}\label{tess_transit_model}

No pre-processed light curves were available for the \emph{TESS} Sector 43 observation of K2-136, so we extracted the photometry from the full frame image (FFI) pixel level. Following \citet{vanderburgetal2019}, we constructed 20 different apertures (10 circular, 10 shaped like the \emph{TESS} point spread function) and selected the one that best minimized photometric scatter. As for \emph{TESS} Sector 44 observations, we used the simple aperture photometry (SAP) light curve produced by the Science Processing Operations Center (SPOC) pipeline \citep{jenkinsetal2016}. Light curves from both sectors were then flattened in the same way: a basis spline fit was performed iteratively on the photometry (with breakpoints every 0.3 days in order to adequately model stellar variability) and $3\sigma$ outliers were removed until convergence (this too, aside from the breakpoint length, follows \citealt{vanderburgetal2019}). Finally, we conducted a simultaneous fit of the low-frequency variability and the transits in order to determine the best-fit low-frequency variability.

\emph{TESS} photometry is not incorporated into our final photometric model, although we did run exploratory joint transit models on \emph{K2} and \emph{TESS} photometry simultaneously. The transit signals of the two smaller planets, K2-136b and d were too small to reliably detect in the \emph{TESS} photometry: individual transits were indistinguishable in depth and quality from temporally adjacent stellar activity. However, the transits of K2-136c were easily identifiable individually in \emph{TESS} photometry, so we ran a joint transit model on all \emph{K2} photometry and \emph{TESS} photometry to explore the resulting improvements in the parameters of K2-136c. This joint transit model included all parameters listed in Section~\ref{transit_model} as well as additional baseline offset and white noise parameters for the two \emph{TESS} Sectors (43 and 44) and two additional quadratic limb-darkening parameters for \emph{TESS} photometry for six additional parameters total. The fit resulted in consistent values for all planet and system parameters as well as a dramatically more precise ephemeris for K2-136c: $P_c = 17.307081^{+0.000014}_{-0.000013}$ days and $t_{0,c} = 8678.07179^{+0.00067}_{-0.00063}$ (BJD-2450000). For comparison, this period and transit time have uncertainties that are both approximately $15$x tighter than those resulting from \emph{K2} transit modeling alone (see Table~\ref{transit_results}). We report these values here to minimize ephemeris drift and facilitate planning of future transit observations of K2-136c.

\subsection{RV Model}\label{rv_model}

We modeled the RV signal of the orbiting planets and the stellar activity simultaneously. We assumed non-interacting planets with Keplerian orbits. We used \texttt{RadVel} \citep{fultonetal2018a} to model the RV signal from each planet with 5 parameters: reference epoch, orbital period, RV semi-amplitude, eccentricity, and argument of periastron. The latter two parameters, eccentricity and longitude of periastron, were parameterized as $\sqrt{e}\cos w$ and $\sqrt{e}\sin w$. As explained in \citet{eastmanetal2013}, this reparameterization avoids a boundary condition at zero eccentricity that may lead to eccentricity estimates that are systematically biased upward. We conducted trial simulations with circular versus eccentric orbits for all three planets and found excellent agreement in all parameters (less than $1\sigma$). We opted to keep eccentricity and argument of the periastron as parameters in order to constrain or place upper limits on each planet's eccentricity. Additionally, we prevented system configurations that would lead to orbit crossings of any two planets, as well as overlaps of planetary Hill spheres. For each planet's reference epoch and orbital period, we applied a Gaussian prior based on the transit parameters determined from M18. We analyzed the \emph{K2} transit photometry with and without \emph{TESS} photometry and verified these M18 values (see Table~\ref{transit_results}).

We also applied additional prior limits on the eccentricities of K2-136b and K2-136d. Based on preliminary modeling of all three planets with uninformed prior eccentricity constraints (and everything else identical to our final model), we found that we could determine the eccentricity of K2-136c but not its siblings. Thus, we decided to set eccentricity constraints by using the Stability of Planetary
Orbital Configurations Klassifier (\texttt{SPOCK}; \citealt{tamayoetal2020}), an N-body simulator that employs machine learning to improve performance. We input into \texttt{SPOCK} our stellar mass posterior and our orbital period posteriors for all three planets (from our preliminary simulations). We also input the mass, eccentricity, and argument of the periastron posteriors for K2-136c (again, from our early simulations). Lastly, we input uniform distributions for mass, eccentricity, and argument of the periastron for K2-136b and K2-136d. Planet mass ranged from 0 up to that of approximately $100\%$ iron planet composition ($3$ M$_\oplus$ for K2-136b and $30$ M$_\oplus$ for K2-136d; \citealt{fortneyetal2007}). Eccentricity ranged from 0 to 0.6 and argument of the periastron ranged from 0 to 360 degrees. We took the subset of our sample that had a $>90\%$ chance of surviving for $>10^9$ orbits (of the innermost planet) and then determined the $3\sigma$ upper limit on the eccentricities of K2-136b and K2-136d for that subset. We then used those values, $e_{\mathrm{b,max}} = 0.35$ and $e_{\mathrm{d,max}} = 0.37$, as eccentricity upper limit priors in all subsequent simulations.

Given the M dwarf companion to our host star \citep{ciardietal2018}, we wanted to include the potential for an RV trend caused by this companion. For further discussion of binarity and linear trends, see Section~\ref{binarity}. Our planetary RV signals and the RV trend therefore take the following form:

\begin{equation} \label{eqn:1}
    RV = \sum_{i \in \{b,c,d\}} K_i\bigg(\cos\big(\omega_i + f_i\big) + e_i\cos\omega_i\bigg) + mt
\end{equation}

\begin{equation} \label{eqn:1b}
    f = 2\arctan\bigg(\sqrt{\dfrac{1+e}{1-e}}\tan\dfrac{E}{2}\bigg)
\end{equation}

\begin{equation} \label{eqn:1c}
    M = E - e\sin E
\end{equation}

\begin{equation} \label{eqn:1d}
    M = nt = \dfrac{2\pi (t-\tau)}{P}
\end{equation}

where $K$ is the induced RV semi-amplitude, $\omega$ is the argument of the periastron, $f$ is the true anomaly, $e$ is the eccentricity, $m$ is the slope of the RV trend, $t$ is the observation time, $E$ is the eccentric anomaly, $M$ is the mean anomaly, $n$ is the mean motion, $\tau$ is the time of periastron passage (as calculated from transit time, orbital period, eccentricity, and argument of the periastron), and $P$ is the orbital period.

The stellar activity was handled via a GP on RVs. We first used a simultaneous model of stellar activity on four different data types: RV, FWHM (a measure of the width of absorption lines), BIS (a measure of line asymmetry), and $S_{HK}$ (an estimate of chromospheric magnetic activity via emission in the cores of the Ca II H \& K lines). However, we found that this approach forced the model to include unreasonable amounts of white noise into each data type via a white noise jitter parameter included in our model. Further, it did not lead to a notable improvement in our final parameter constraints. We conducted numerous tests exploring the excess white noise preferred by the model, including modeling different instruments, different numbers of planets, altering the data reduction pipeline (e.g. trying the ZLSD pipeline; \citealt{lienhardetal2022}), and creating synthetic data sets to model against. The most reasonable hypothesis we could find is that for the K2-136, our data (especially for ESPRESSO) is of a sufficiently high quality, with small enough uncertainties, that the model we used from \citet{rajpauletal2015} to relate the stellar activity indices to each other and to the RVs was not complex enough to account for the correlated structure of the stellar activity of K2-136.

\begin{figure*}[ht!]
\epsscale{1.15}
  \begin{center}
      \leavevmode
\plotone{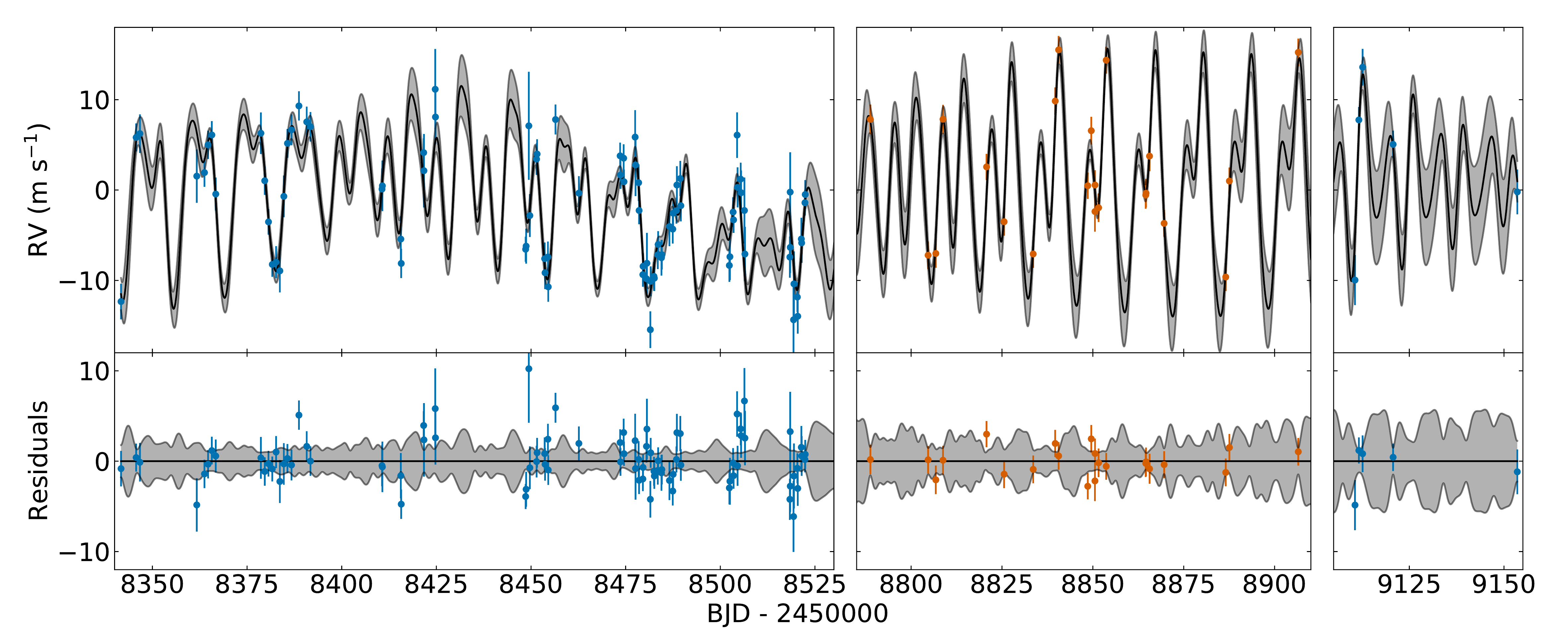}
\caption{K2-136 observations and model fits for RVs (and fit residuals). In each panel, the blue points (HARPS-N) or orange points (ESPRESSO) are the observations while the black line and gray region are the model fit and $1\sigma$ confidence interval, respectively. RVs have been mean-subtracted (corresponding to their respective instrument) and planet-induced reflex motion has been subtracted as well. RV errors have been inflated from their original values by adding the model-estimated RV jitter term in quadrature. Note the two time gaps between the first 88 HARPS-N observations, the 22 ESPRESSO observations, and the final 5 HARPS-N observations.} \label{stellar_activity}
\end{center}
\end{figure*}

Our RV data and the final RV model fit can be seen in Fig.~\ref{stellar_activity}. Despite only modeling RVs without any stellar activity indices, a GP is still robust and allows us to separate planet-induced RVs from stellar activity to the extent that disentanglement is possible. We followed the method laid out in \citet{rajpauletal2015}, but we constrained the model only to the portions relevant to RVs rather than additional stellar activity indices. They dictate that the RVs are related to the stellar activity as follows:

\begin{equation} \label{eqn:2}
    \Delta RV = V_cG(t) + V_r\dot{G}(t)
\end{equation}

In this equation, $G(t)$ corresponds to the underlying stellar activity GP, while $V_c$ and $V_r$ correspond to the RV amplitudes of the convective blueshift and rotation modulation effect, respectively. It is important to include both rotation modulation and convective blueshift for data types impacted by both, as one phenomenon may play a larger role than the other depending on the data type and star. In fact, our final results (see Table~\ref{rv_results}) show that for K2-136, rotation modulation has an outsized effect on RVs compared to convective blueshift; however, we retain both terms in our model since $V_c$ is still inconsistent with zero. We follow \citet{rajpauletal2015} further by using a quasi-periodic kernel to establish the covariance matrix of our GP. A quasi-periodic kernel is a good choice to capture the stellar variability of a star because quasi-periodicity describes well the variability exhibited by a rotating star with starspots that come and go. Following \citet{gilesetal2017}, we find a predicted starspot lifetime for K2-136 of $38^{+20}_{-13}$ days (versus a rotation period of $13.37$ days); this estimate is consistent with our GP evolution time-scale result ($48^{+19}_{-11}$ days, see Table~\ref{rv_results}). In other words, stellar activity during the time frame of a single rotation period is likely to look similar (though not identical) to stellar activity during the previous and subsequent stellar rotation periods, since the starspots for K2-136 are likely to be longer lived than the stellar rotation period. Thus, the quasi-periodic kernel is defined as follows:

\begin{multline} \label{eqn:QP}
    K_{\mathrm{QP}}(t_i,t_j) = \\ h^2\exp{\bigg(-\dfrac{\sin^2{\big(\pi(t_i-t_j)/P_*\big)}}{2\lambda^2_p}-\dfrac{(t_i-t_j)^2}{2\lambda_e^2}\bigg),}
\end{multline}

where $h$ is the GP amplitude (which is folded into the amplitude parameters described above), $P_*$ is the stellar stellar rotation period, $\lambda_e$ is a decay timescale proportional to the starspot lifetime, and $\lambda_p$ is a smoothness parameter that captures the level of variability within a single rotation period. $t_i$ and $t_j$ are any two times between which the covariance is being calculated; for a given time series of $N$ observations, all $N^2$ combinations of time pairs create the $N$x$N$ covariance matrix. This covariance matrix (plus a mean model) is a normal multivariate distribution; $G(t)$ can be explored by sampling from this distribution. We refer the reader to \citet{rajpauletal2015} for a more detailed description of this method, as well as \citet{mayoetal2019} for an application of this method to the sub-Neptune Kepler-538b. 

Finally, we include a jitter parameter (added in quadrature to RV uncertainties) and a baseline offset parameter for each instrument (HARPS-N and ESPRESSO). All together, our full model includes 21 planetary parameters (5 per planet: time of transit, orbital period, RV semi-amplitude, $\sqrt{e}\sin{\omega}$, and $\sqrt{e}\cos{\omega}$), 5 GP parameters (2 GP amplitudes corresponding to $V_c$ and $V_r$ from equation~\ref{eqn:2} as well as $P_*$, $\lambda_p$, and $\lambda_e$ from equation~\ref{eqn:QP}), 1 linear trend parameter, 2 RV noise parameters (1 per instrument) and 2 RV baseline offset parameters (1 per instrument), for a total of 25 parameters.

We used a uniform prior for the RV semi-amplitude for each planet. However, we conducted a trial simulation to compare a uniform versus log uniform prior for RV semi-amplitude and found no discernible difference in our results (all parameters agreed to within $1\sigma$).

\begin{figure*}[ht!]
\epsscale{1.15}
  \begin{center}
      \leavevmode
\plotone{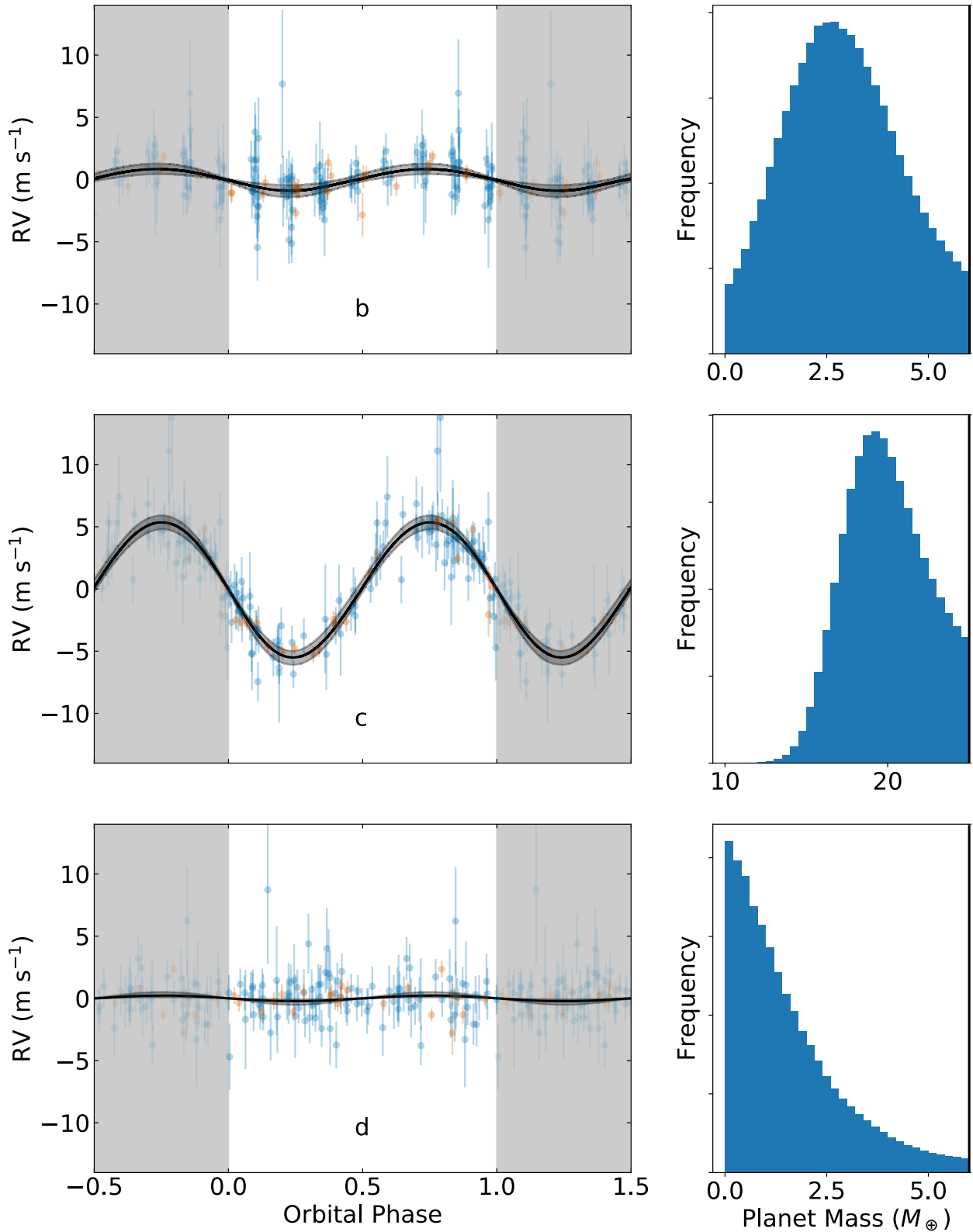}
\caption{Left: Phased RV plots for all three K2-136 planets. For each subplot, we used our best fit model parameters to remove stellar activity and the presence of the other two planets. In each subplot, blue data points (HARPS-N) and orange data points (ESPRESSO) are unbinned RV observations. The black line is the median fit and the gray region around that line is the $1\sigma$ confidence interval. Right: Posterior mass distribution plots for all three K2-136 planets. It is visually apparent that while there is a strong mass detection for K2-136c ($R_p = 3.00\pm0.13$ R$_\oplus$), there is at best only a marginal detection for K2-136b ($R_p = 1.014\pm0.050$ R$_\oplus$) and no evidence of a detection for K2-136d ($R_p = 1.565\pm0.077$ R$_\oplus$). Therefore, therefore we report only place upper limits on the masses of K2-136b and K2-136d (see Table~\ref{rv_results}).} \label{phase_plot}
\end{center}
\end{figure*}

\begin{figure*}
    \adjustimage{width=.65\textwidth,center}{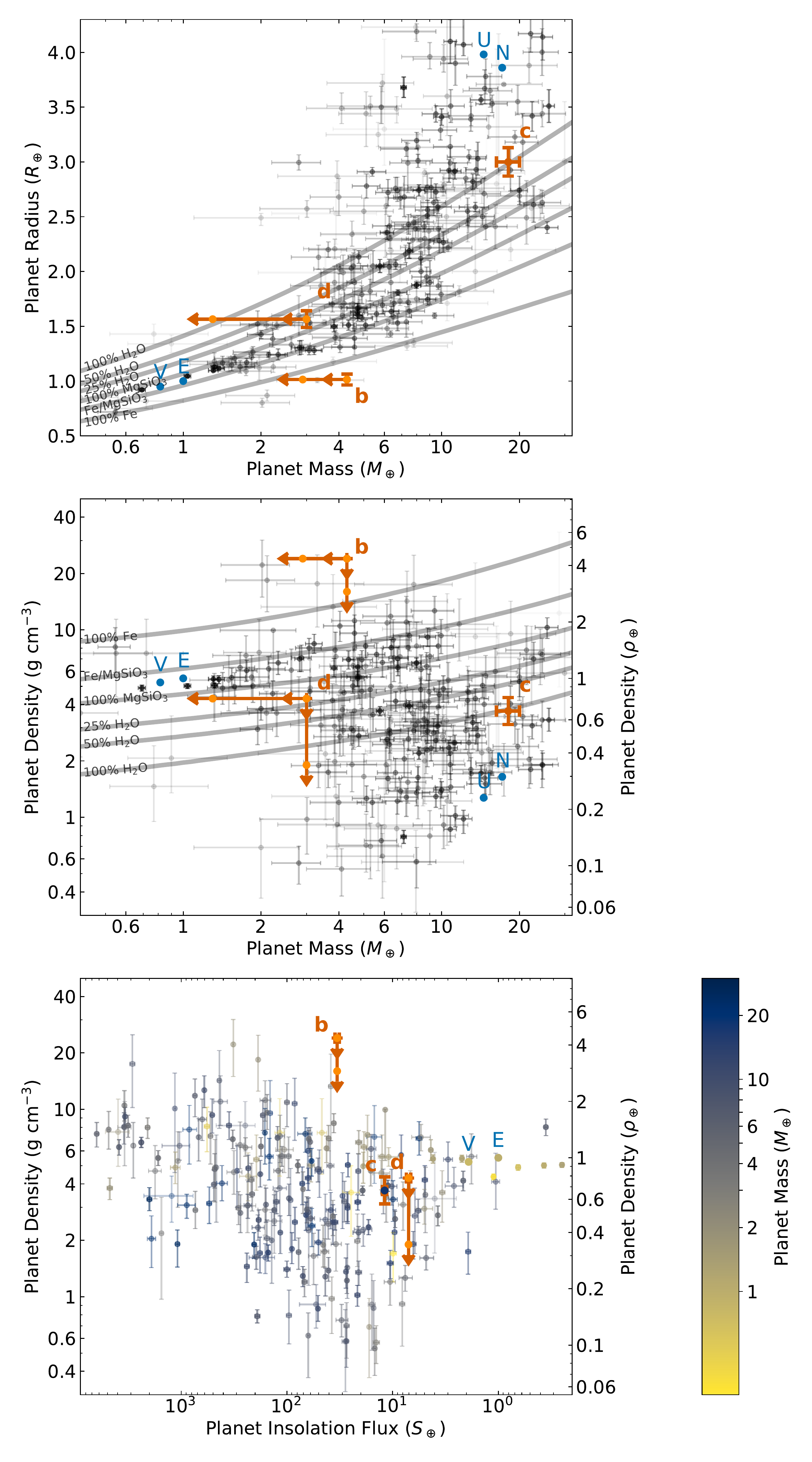}
    \caption{Top: mass-radius diagram of transiting planets with fractional mass and radius uncertainties less than 50\%. K2-136b, c, and d are plotted in dark orange, with mass uncertainties on K2-136b and d as 68\% and 95\% upper limits denoted in light orange. 
    Data collected from the NASA Exoplanet Archive (accessed 2023 Mar 22; \citealt{https://doi.org/10.26133/nea12}). Venus, Earth, Uranus, and Neptune are also labeled and plotted in blue for reference. Except for the K2-136 system, planets with larger fractional mass and radius uncertainties are fainter. Gray lines correspond to planetary compositions (from top to bottom) of 100\% H$_2$O, 50\% H$_2$O, 25\% H$_2$O, 100\% MgSiO$_3$, 50\% MgSiO$_3$ + 50\% Fe, and 100\% Fe, respectively \citep{zengandsasselov2013,zengetal2016}. Kepler-136c lies closest to the 100\% H$_2$O composition line, and is similar in mass to Uranus and Neptune although smaller and much more dense. Middle: the same sample plotted in mass versus planet density, with the same solar system references and composition lines (order inverted from top panel). Bottom: the same sample plotted in planet insolation flux versus planet density, with the color of all data points corresponding to planet mass (except K2-136b and d, which only have mass upper limits and thus are plotted in light orange).} \label{mr_plot}
\end{figure*}

\begin{table*}[t]
\begin{center}
\caption{K2-136 transit and planetary parameters \label{transit_results}}
\begin{tabular}{llcc}
\tableline
\tableline
Parameter & Unit & This Paper & Priors \\
\tableline
\multicolumn{4}{l}{\textit{Planet b}} \\
Period $P_b$ & day & $7.97525 \pm 0.00073$ & Unif($7.96529$, $7.98529$) \\
Time of transit $t_{0,b}$ & BJD-2450000 & $8679.083^{+0.075}_{-0.074}$ & Unif($8678.58762$, $8679.58762$)\footnote[1]{$t_{0}$ centered between \emph{K2} and \emph{TESS} photometry, so ephemeris drift is incorporated into $t_{0}$ and $P$.} \\
Planet-star radius ratio $R_b/R_*$ & - & $0.01370^{+0.00041}_{-0.00036}$ & Jeffreys($0.001$, $0.1$) \\
Radius $R_b$ & R$_\oplus$ & $1.014^{+0.050}_{-0.049}$ & ... \\
Transit duration $T_{14,b}$ & hr & $2.67^{+0.086}_{-0.084}$ & Unif($0$, $7.2$) \\
Impact parameter $b_b$ & - & $0.22^{+0.15}_{-0.14}$ & Unif($0$, $1$) \\
Semi-major axis $a_b$ & AU & $0.0707 \pm 0.0012$ & ... \\
Mean density $\rho_b$ & $\rho_\oplus$ & $<2.8$\footnotemark[2], $<4.4$\footnotemark[3] & ... \\
Mean density $\rho_b$ & g cm$^{-3}$ & $<16$\footnotemark[2], $<24$\footnotemark[3] & ... \\
Insolation flux $S_b$ & S$_\oplus$ & $33.5^{+1.6}_{-1.5}$ & ... \\
Equilibrium temperature $T_{eq,b}$ (albedo $= 0.3$) & K & $610$ & ... \\
Equilibrium temperature $T_{eq,b}$ (albedo $= 0.5$) & K & $560$ & ... \\\\

\tableline
\multicolumn{4}{l}{\textit{Planet c}} \\

Period $P_c$ & day & $17.30723^{+0.00019}_{-0.00020}$ & Unif($17.30514$, $17.30914$) \\
Time of transit $t_{0,c}$ & BJD-2450000 & $8678.0792^{+0.0088}_{-0.0096}$ & Unif($8677.0747$, $8679.0747$)\footnotemark[1] \\
Planet-star radius ratio $R_c/R_*$ & - & $0.04064^{+0.00068}_{-0.00071}$ & Jeffreys($0.001$, $0.1$) \\
Radius $R_c$ & R$_\oplus$ & $3.00 \pm 0.13$ & ... \\
Transit duration $T_{14,c}$ & hr & $3.449^{+0.039}_{-0.031}$ & Unif($0$, $7.2$) \\
Impact parameter $b_c$ & - & $0.31^{+0.11}_{-0.14}$ & Unif($0$, $1$) \\
Semi-major axis $a_c$ & AU & $0.1185^{+0.0020}_{-0.0021}$ & ... \\
Mean density $\rho_c$ & $\rho_\oplus$ & $0.67^{+0.12}_{-0.10}$ & ... \\
Mean density $\rho_c$ & g cm$^{-3}$ & $3.69^{+0.67}_{-0.56}$ & ... \\
Insolation flux $S_c$ & S$_\oplus$ & $11.91^{+0.57}_{-0.53}$ & ... \\
Equilibrium temperature $T_{eq,c}$ (albedo $= 0.3$) & K & $470$ & ... \\
Equilibrium temperature $T_{eq,c}$ (albedo $= 0.5$) & K & $440$ & ... \\\\

\tableline
\multicolumn{4}{l}{\textit{Planet d}} \\

Period $P_d$ & day & $25.5750^{+0.0022}_{-0.0021}$ & Unif($25.5551$,$25.5951$) \\
Time of transit $t_{0,d}$ & BJD-2450000 & $8675.936^{+0.072}_{-0.068}$ & Unif($8675.4401$, $8676.4401$)\footnotemark[1] \\
Planet-star radius ratio $R_d/R_*$ & - & $0.02119^{+0.00057}_{-0.00061}$ & Jeffreys($0.001$, $0.1$) \\
Radius $R_d$ & R$_\oplus$ & $1.565^{+0.077}_{-0.076}$ & ... \\
Transit duration $T_{14,d}$ & hr & $3.04^{+0.10}_{-0.09}$ & Unif($0$, $7.2$) \\
Impact parameter $b_d$ & - & $0.677^{+0.042}_{-0.049}$ & Unif($0$, $1$) \\
Semi-major axis $a_d$ & AU & $0.1538^{+0.0026}_{-0.0027}$ & ... \\
Mean density $\rho_d$ & $\rho_\oplus$ & $<0.35$\footnotemark[2], $<0.79$\footnotemark[3] & ... \\
Mean density $\rho_d$ & g cm$^{-3}$ & $<1.9$\footnotemark[2], $<4.3$\footnotemark[3] & ... \\
Insolation flux $S_d$ & S$_\oplus$ & $7.07^{-0.34}_{-0.32}$ & ... \\
Equilibrium temperature $T_{eq,d}$ (albedo $= 0.3$) & K & $420$ & ... \\
Equilibrium temperature $T_{eq,d}$ (albedo $= 0.5$) & K & $380$ & ... \\\\

\tableline
\multicolumn{4}{l}{\textit{System parameters}} \\

\Kepler/\emph{K2} quadratic limb-darkening parameter $q_{1,Kepler}$ & - & $0.38^{+0.22}_{-0.13}$ & Unif($0$, $1$) \\
\Kepler/\emph{K2} quadratic limb-darkening parameter $q_{2,Kepler}$ & - & $0.56^{+0.24}_{-0.18}$ & Unif($0$, $1$) \\
\emph{K2} Campaign 13 normalized baseline offset & ppm & $0.7^{+2.9}_{-3.0}$ & Unif($-1000$, $1000$) \\
\emph{K2} Campaign 13 photometric white noise amplitude & ppm & $53.9 \pm 1.8$ & ModJeffreys($1$, $1000$, $0$) \\

\tableline
\end{tabular}
\end{center}
\end{table*}

\begin{table*}[t]
\begin{center}
\caption{K2-136 RV model parameters \label{rv_results}}
\begin{tabular}{llcc}
\tableline
\tableline
Parameter & Unit & This Paper & Priors \\
\tableline
\multicolumn{4}{l}{\textit{Planet b}} 
\\

Period $P_b$ & day & $7.97520 \pm 0.00079$ & Normal($7.97529$, $0.00080$)\footnote[1]{\citet{mannetal2018}} \\
Time of transit $t_{0,b}$ & BJD-2450000 & $7817.7563^{+0.0046}_{-0.0048}$ & Normal($7817.7563$, $0.0048$)\footnotemark[1] \\
Semi-amplitude $K_b$ & m s$^{-1}$ & $<1.2$\footnote[2]{$68\%$ confidence limit}, $<1.7$\footnote[3]{$95\%$ confidence limit} & Unif($0.001$, $20$) \\
Eccentricity $e_b$ & - & $0.14^{+0.12}_{-0.11}$ ($<0.21$\footnotemark[2], $<0.32$\footnotemark[3]) & d\footnotetext[4]{Unif($-1$, $1$) on $\sqrt{e}\sin{\omega}$ and $\sqrt{e}\cos{\omega}$. See \citet{eastmanetal2013}.}, e\footnotetext[5]{K2-136b and K2-136d had additional eccentricity prior upper limits of $0.35$ and $0.37$, respectively; see Section~\ref{rv_model}.} \\
Argument of periastron $\omega_b$ & degrees & $189^{+82}_{-132}$ & d, e \\
Mass $M_b$ & M$_\oplus$ & $<2.9$\footnotemark[2], $<4.3$\footnotemark[3] & ... \\\\

\tableline
\multicolumn{4}{l}{\textit{Planet c}} \\

Period $P_c$ & day & $17.30713 \pm 0.00027$ & Normal($17.30714$, $0.00027$)\footnotemark[1] \\
Time of transit $t_{0,c}$ & BJD-2450000 & $7812.71770^{+0.00086}_{-0.00085}$ & Normal($7812.71770$, $0.00089$)\footnotemark[1] \\
Semi-amplitude $K_c$ & m s$^{-1}$ & $5.49^{+0.54}_{-0.52}$ & Unif($0.001$, $20$) \\
Eccentricity $e_c$ & - & $0.047^{+0.062}_{-0.034}$ ($<0.074$\footnotemark[2], $<0.16$\footnotemark[3]) & d \\
Argument of periastron $\omega_c$ & degrees & $124 \pm 99$ & d \\
Mass $M_c$ & M$_\oplus$ & $18.1^{+1.9}_{-1.8}$ & ... \\\\

\tableline
\multicolumn{4}{l}{\textit{Planet d}} \\

Period $P_d$ & day & $25.5750^{+0.0024}_{-0.0023}$ & Normal($25.5751$, $0.0024$)\footnotemark[1] \\
Time of transit $t_{0,d}$ & BJD-2450000 & $7780.8117 \pm 0.0065$ & Normal($7780.8116$, $0.0065$)\footnotemark[1] \\
Semi-amplitude $K_d$ & m s$^{-1}$ & $<0.36$\footnotemark[2], $<0.78$\footnotemark[3] & Unif($0.001$, $20$) \\
Eccentricity $e_d$ & - & $0.071^{+0.063}_{-0.049}$ ($<0.10$\footnotemark[2], $<0.16$\footnotemark[3]) & d, e \\
Argument of periastron $\omega_d$ & degrees & $280^{+130}_{-110}$ & d, e \\
Mass $M_d$ & M$_\oplus$ & $<1.3$\footnotemark[2], $<3.0$\footnotemark[3] & ... \\\\

\tableline
\multicolumn{4}{l}{\textit{System and GP parameters}} \\

RV slope & m s$^{-1}$ yr$^{-1}$ & $0.1 \pm 2.5$ & Unif($-365$, $365$) \\
HARPS-N RV white noise amplitude & m s$^{-1}$ & $0.83 \pm 0.52$ & Unif($0$, $20$) \\
ESPRESSO RV white noise amplitude & m s$^{-1}$ & $1.57^{+0.73}_{-0.62}$ & Unif($0$, $20$) \\
HARPS-N RV offset amplitude & m s$^{-1}$ & $39503.7^{+2.1}_{-1.9}$ & Unif($39450$, $39550$) \\
ESPRESSO RV offset amplitude & m s$^{-1}$ & $39494.7^{+3.5}_{-3.6}$ & Unif($39450$, $39550$) \\
GP RV convective blueshift amplitude $V_c$ & m s$^{-1}$ & $3.5^{+2.2}_{-1.2}$ & Unif($0$, $100$) \\
GP RV rotation modulation amplitude $V_r$ & m s$^{-1}$ & $22.0^{+12.9}_{-8.0}$ & Unif($0$, $100$) \\
GP stellar rotation period $P_*$ & day & $13.37^{+0.13}_{-0.17}$ & Unif($1$, $20$) \\
GP inverse harmonic complexity $\lambda_p$ & ... & $0.75^{+0.23}_{-0.16}$ & Unif($0.1$, $3$) \\
GP evolution time-scale $\lambda_e$ & day & $48^{+19}_{-11}$ & Unif($1$, $200$) \\

\tableline
\end{tabular}
\end{center}
\end{table*}

\begin{table*}[htp]
\begin{center}
\caption{Model Evidence Comparisons for K2-136 Planet Configurations}\label{evidence_table}
\begin{tabular}{ |c|c|c| } 
 \hline
 Planets in model & $\Delta\log_{10}$(evidence) & Interpretation\\ 
 \hline
 c & 0 & - \\ 
 \hline
 b,c & -1.10 & Strongly disfavored \\ 
 \hline
 c,d & -2.06 & Strongly disfavored \\ 
 \hline
 b,c,d & -3.23 & Excluded \\ 
 \hline
 - & -5.54 & Excluded \\ 
 \hline
 d & -7.00 & Excluded \\ 
 \hline
 b & -7.78 & Excluded \\ 
 \hline
 b,d & -9.17 & Excluded \\ 
 \hline
\end{tabular}
\end{center}
\end{table*}

\subsection{Parameter estimation}\label{parameter_estimation}

We conducted parameter estimation of our model with the observed data using the Bayesian inference tool \texttt{MultiNest} \citep{ferozetal2009,ferozetal2013}. We set \texttt{MultiNest} to constant efficiency mode, importance nested sampling mode, and multimodal mode. We used a sampling efficiency of 0.01, 1000 live points, and an evidence tolerance of 0.1. Constant efficiency is typically off, but we turned it on since it allows for better exploration of parameter space in higher-dimensional models such as our own. Further, sampling efficiency is usually set to 0.8 and the number of live points is usually set to 400. Decreasing sampling efficiency and increasing the number of live points leads to more complete coverage of parameter space, at the cost of a typically longer simulation convergence time. Finally, the evidence tolerance is usually set to 0.8; reducing the evidence tolerance causes the simulation to run longer but increases confidence that the simulation has fully converged. In other words, the standard \texttt{MultiNest} settings would likely lead to reliable results, but our choice of settings increases the trustworthiness of our parameter estimation and model evidence results.

\subsection{Model Evidence Comparison}\label{evidences}

One of the strengths of \texttt{MultiNest} is that it automatically calculates the Bayesian evidence of the selected model, making model comparison very easy. We compared the model evidences of eight different models (RV only, no photometry), based on all possible combinations of planets b, c, and d. The results of our model comparisons are listed in Table~\ref{evidence_table}. We find that the most preferred model is the one that contains only planet c, and not planets b or d. In fact, using the Bayes factor interpretation of \citet{kassandraftery1995}, we find that almost every other combinations of planets can be decisively ruled out (i.e. the Bayes factor of the planet c model to any other model is $>100$). The only exceptions are the model with planets b and c and the model with planets c and d, which are only strongly disfavored. This tells us that neither K2-136b nor K2-136d are unambiguously detected, so including either in the model quickly worsens the model evidence. However, it is notable that the model with planets b and c is better than the model with planets c and d, and is nearly in the more likely ``Disfavored'' category rather than ``Strongly disfavored''. This makes sense, as K2-136b (unlike K2-136d) has a non-zero peak in its semi-amplitude posterior distribution (see Fig.~\ref{phase_plot}). In fact, although only upper limits are reported for the mass of K2-136b in Table~\ref{rv_results}, the median mass is actually non-zero at the $2.0\sigma$ level ($M_b$ = $2.4 \pm 1.2$ M$_\oplus$).

Our results tell us that we can be confident that K2-136c has been detected in our observations. In contrast, the RV signals of K2-136b and K2-136d fall below the threshold of detection, at least given the quantity and quality of our specific data set. Continued radial velocity monitoring, particularly with high precision instruments and facilities may be able to measure their masses, especially K2-136b (which appears to already be near the threshold of detection with our current data set).

Although the model with only K2-136c is the most favored, we still use a model with all three planets as our canonical model for parameter estimation for three reasons. First, we already know from transit photometry that planets b and d exist. Accordingly, the goal of the model comparison exercise described above is not to question the existence of these planets but to examine whether the RV signals from each planet can be detected in our data set. By adopting the three-planet model, we incorporate the uncertainties introduced by the unknown masses and eccentricities of planets b and d. Second, using the three-planet model allows us to determine upper limits for the masses of planets b and d, which is useful for constraining planet compositions and providing guidance for any future attempts to constrain the mass of either planet. Third, both models agree very closely: all parameters are consistent at $1\sigma$ or less, and all uncertainties on parameters are similar in scale. For the RV semi-amplitude of K2-136c, our key parameter of interest, our canonical model returned $K_c = 5.49^{+0.54}_{-0.52}$ m s$^{-1}$ while the one-planet model returned $K_c = 5.17^{+0.56}_{-0.51}$ m s$^{-1}$.

\subsection{Model Reliability Tests}

In order to rigorously assess the accuracy of our results, we conducted tests to analyze different components of our model. Specifically, we removed the GP portion of the model, and we also injected and recovered synthetic planets into the system to compare input and output RV semi-amplitudes.

For our test models we chose not to include planets b and d, as well as photometry, and we then compared against the model with only K2-136c (hereafter referred to as the ``one-planet reference model'') rather than the three-planet model; this is despite already selecting the three-planet model as the canonical model to report our results (see Section~\ref{evidences}). This was done for a few reasons. First, the one-planet model is the preferred model according to the Bayesian evidences, so it is a reasonable point of comparison. Second, as stated earlier, the one-planet reference model and the three-planet model agree very closely. Therefore, any test model parameters found to be highly consistent with the one-planet reference model results will also be highly consistent with the three-planet model results. Third, as a practical matter, including only K2-136c in our test models significantly reduced computational complexity, allowing us to test a wider variety of models.

\subsubsection{No GP} \label{noGP}
GPs are very versatile and can fit highly variable and correlated signals. Therefore, it is reasonable to ask whether a GP may, in the process of modeling stellar activity, ``steal'' part of the RV signal from a planet due to overfitting of the data. In order to address these concerns, we ran a model without a GP, and no alternative method to handle stellar activity. We found the resulting parameters were broadly consistent. The noise parameters for each data type in the no-GP model were notably larger, but that is to be expected given no mitigation of the stellar activity. All other parameters agreed with the one-planet reference model parameters to within $1\sigma$; for the RV semi-amplitude of K2-136c, we determined a value of $K_c = 5.92^{+0.89}_{-0.91}$ m s$^{-1}$ (compared to $K_c = 5.17^{+0.56}_{-0.51}$ m s$^{-1}$ for the one-planet reference model). Finally, it is worth noting we also found that $\Delta \log_{10}$(evidence) $=-18.1$ compared to the one-planet reference model, decisively ruling out the no-GP model \citep{kassandraftery1995}. In other words, a GP accounts for the stellar activity satisfactorily, whereas ignoring stellar activity is clearly insufficient.

\subsubsection{Synthetic Planet Injections}

We also conducted planet injection tests to determine how robustly we could recover the injected signals. Accurate recovery of such signals builds confidence in the accuracy of the RV signal recovered for K2-136c as well as the upper limits placed on K2-136b and K2-136d.

We ran four separate tests in which we injected a $5.5$ m s$^{-1}$ RV signal of a planet on a circular orbit with a period of 4, 12, 20, and 28 days. $5.5$ m s$^{-1}$ was chosen because it is approximately the same semi-amplitude as the signal induced by K2-136c, allowing us to directly test our confidence in the recovered RV signal of K2-136c specifically. Our set of orbital periods was selected in order to 1) span the range of known periodic signals in the system (the orbital period of the three known planets and the stellar rotation period), 2) avoid close proximity to those signals (none are within 1.5 days of the injected signals), and 3) be equally spaced in order to uniformly test the encompassed period range.

All four of the recovered signals agree with the injected signal of $5.5$ m s$^{-1}$ to within $1\sigma$. In all four tests, the recovered signal of K2-136c also agrees with our one-planet reference model ($K_c = 5.28 \pm 0.56$ m s$^{-1}$) within $1\sigma$, lending further confidence to our results.

\subsection{High-Energy Observations} \label{HST_discussion}

\subsubsection{\textit{HST }NUV observations}

To compare the UV quiescent activity of K2-136 with other K stars Hyades members, we measured the surface flux of the Mg II h ($2796.35$ \AA) and k ($2808.53$ \AA) lines. The Mg II lines are the strongest emission lines in the NUV and correlate strongly with the chromospheric activity of the star. For accurate emission measurements, we subtracted the NUV continuum by fitting the data outside of the Mg II integration region using the \texttt{astropy} module \texttt{specutils}. We then integrated over $2792.0 - 2807.0$~\AA\ to measure the Mg II emission flux. To convert from observed flux to surface flux, we estimated the radii of the stars using the relationships between age, effective temperature, and radius given by \citet{baraffeetal2015}, except for K2-136, where we use the radius reported in this work.

\begin{figure}[t!]
\epsscale{1.1}
  \begin{center}
      \leavevmode
\plotone{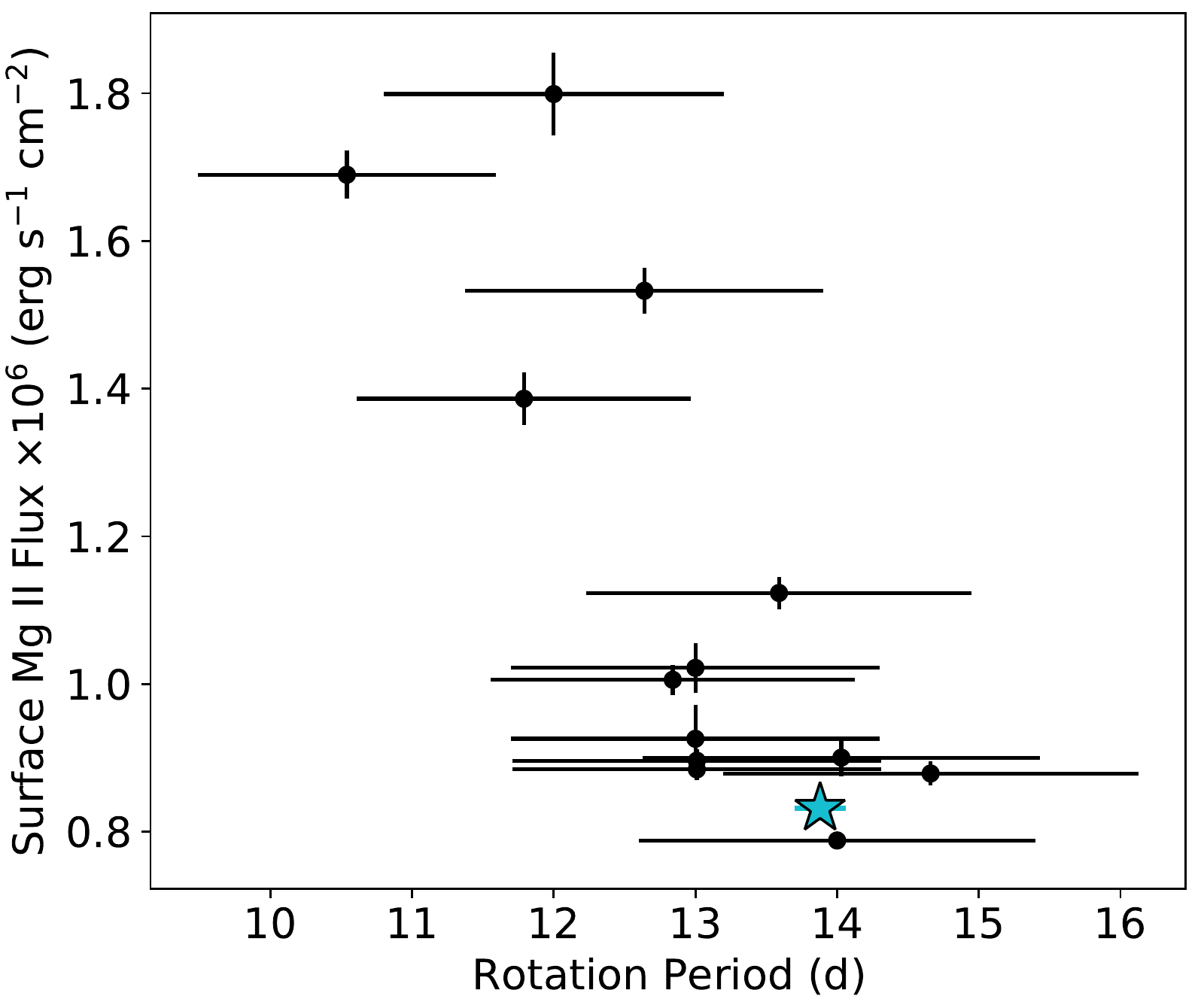}
\caption{Mg II surface flux as a function of stellar rotation period for K star Hyades members. The blue star represents K2-136, and the black points are the 13 other K star Hyades observed in a broader HST program (GO-15091, PI: Ag\"ueros). The rotation periods are from \citet{Douglas2019} and have assumed errors of 10\%, except for K2-136, which we determined to be $13.88^{+0.17}_{-0.18}$~d in this work. K2-136 does not show any distinct chromospheric activity or unique rotation period compared to the rest of the sample.} \label{fig:uv-activity}
\end{center}
\end{figure}

Figure \ref{fig:uv-activity} shows the Mg II surface flux as a function of the rotation period for K2-136 and 13 other observed K star members of the Hyades from GO-15091. K2-136 has a surface Mg~II flux of ($8.32\pm0.17$)~$\times$~$10^5$~erg~s$^{-1}$~cm$^{-2}$, whereas the median of the sample is ($9.66\pm0.24$)~$\times$~$10^5$~erg~s$^{-1}$~cm$^{-2}$.

Additionally, \citet{richey-yowelletal2019} measured the NUV flux densities of 97 K stars in the Hyades using archival data from the \textit{Galaxy Evolution Explorer} \citep[{\it GALEX};][]{galex2005}. The median NUV flux density of their sample of Hyades stars at a normalized distance of $10$ pc was $1.89 \times 10^3 {\mu}$Jy. We measure a {\it GALEX} NUV magnitude for K2-136 of $19.51$ mag, which corresponds to a flux density at $10$ pc of $1.52 \times 10^3 {\mu}$Jy, well within the interquartiles of the total sample from \citet{richey-yowelletal2019}.

\begin{figure*}
\epsscale{1.0}
  \begin{center}
      \leavevmode
\plotone{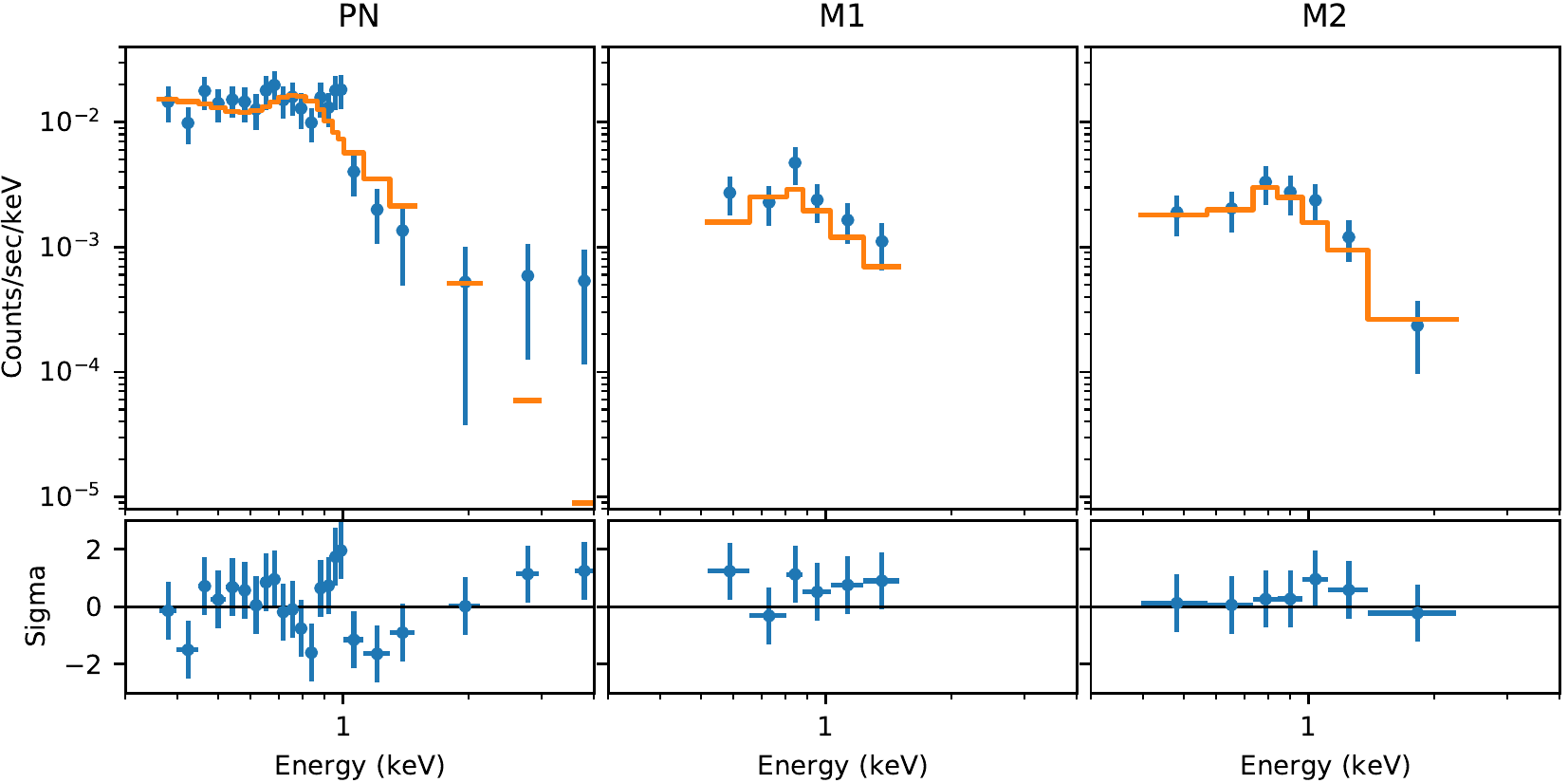}
\caption{X-ray spectra of K2-136 from the XMM pn (left panel), MOS1 (center), and MOS2 (right) EPIC cameras. X-ray counts are binned by 20 in the pn camera, and 15 in the MOS cameras. The orange lines are the best fits using a one-temperature APEC model and assuming a fixed neutral hydrogen column density of $5.5\times10^{18}$~cm$^{-2}$, typical for Hyads (see Sec.~\ref{xrays}). The residuals of each fit are shown in the bottom panels. The best-fit parameters are presented in Table~\ref{xrays_table}.} \label{fig:xrayspec}
\end{center}
\end{figure*}

\subsubsection{X-ray observations}\label{xrays}
We detected a total of 800 EPIC counts for K2-136 in the XMM observation and can therefore extract an X-ray spectrum for the star. We used a one-temperature APEC model to fit the spectrum, which is appropriate for representing the hot plasma in stellar coronae. We combined this model with the ISM absorption model \texttt{tbabs} using photoelectric cross-section from \citet{Balucinska-Church1992} to account for the neutral hydrogen column density $N_\mathrm{H}$. We set $N_\mathrm{H}$ to $5.5\times10^{18}$~cm$^{-2}$, derived using $E$($B-V$) = 0.001 for Hyades \citep{Taylor2006}, $R_V = 3.1$, and the relation $N_{\mathrm{H}}$[cm$^{-2}/A_v$] = $1.79\times 10^{21}$ \citep{Predehl1995}; allowing $N_\mathrm{H}$ to float did not improve the fit. The spectra for each of the three XMM cameras are shown in Figure~\ref{fig:xrayspec} and the best-fit parameters, which are obtained from a simultaneous fit to the three, are provided in Table~\ref{xrays_table}.

Using the total EPIC energy flux from the spectral best fit, we obtained an X-ray luminosity $L_\mathrm{X}$ = ($1.26 \pm 0.19$)$\times10^{28}$ erg s$^{-1}$ and $L_\mathrm{X}/L_*$ = ($1.97 \pm 0.30$)$\times10^{-5}$ (0.1$-$2.4 keV energy range). These values are within $1\sigma$ of those found by \citet{fernandezfernandezetal2021} using the same XMM observation\footnote[4]{These authors performed spectral fitting (using a three-temperature APEC model) only to the EPIC pn detection to derive the X-ray energy flux for K2-136.}. For the sample of 89 K dwarfs with X-ray detections in the Hyades, the median values for $L_\mathrm{X}$ and $L_\mathrm{X}/L_*$ are $4.5_{-3.3}^{+7.9} \times 10^{28}$ erg s$^{-1}$ and $4.0_{-1.6}^{+20.8} \times 10^{-5}$, respectively (N\'u\~nez et al.~in prep.). K2-136, therefore, appears somewhat less luminous in X-rays than most of its coeval K dwarf brethren. A narrower comparison, against late-K (K5 and later) dwarf Hyads, shows that the $L_\mathrm{X}$ and $L_\mathrm{X}/L_*$ values for K2-136 are within one standard deviation of the median for that cohort.

In addition to X-ray luminosity, we also estimated extreme ultraviolet (EUV) luminosity using stellar age and Equation 4 from \citet{sanz-forcadaetal2011} and found $L_\mathrm{EUV}$ = ($22.6^{+7.8}_{-5.7}$)$\times10^{28}$ erg s$^{-1}$. Combining $L_\mathrm{X}$ and $L_\mathrm{EUV}$ and using the semi-major axis of K2-136c, we are able to estimate the X-ray and UV flux incident on K2-136c to be $F_\mathrm{XUV}$ = ($6.0^{+2.0}_{-1.4}$)$\times10^{3}$ erg s$^{-1}$ cm$^{-2}$. Then, using this incident flux value (along with M$_\mathrm{c}$ and R$_\mathrm{c}$) we estimate an atmospheric mass loss rate with Equation 1 from \citet{fosteretal2022}. This yields a current atmospheric mass loss rate for K2-136c of $\dot{M_\mathrm{c}}$ = ($3.4^{+1.3}_{-0.9}$)$\times10^{9}$ g s$^{-1}$ = ($17.8^{+6.7}_{-5.0}$)$\times10^{-3}$ M$_\oplus$ Gyr$^{-1}$. This rate is based on current values, so the mass loss rate in the past or future may differ. At this current rate, with a H$_2$-He envelope mass fraction of $\sim5\%$, it would take $51^{+23}_{-16}$ Gyr to fully evaporate the atmosphere. Even the $95\%$ lower limit evaporation time is still $28$ Gyr, longer than the age of the universe. In other words, in $\sim4$ Gyr, when the K2-136 system is as old as the Solar System currently is, we expect K2-136c will have likely only lost $5-10\%$ of its current atmosphere.

We also calculated the Rossby number $R_o$ of K2-136, which is defined as the star's rotation period $P_*$ divided by the convective turnover time $\tau$. We used the ($V-K_s$)-log $\tau$ empirical relation in \cite{wrightetal2018} (their equation 5) to obtain $\tau = 22.3$ d for K2-136. Using our measured $P_*$ value (see Sec.~\ref{rotation_period}) gives $R_o = 0.6$. This $R_o$ puts K2-136 well within the X-ray unsaturated regime, in which the level of magnetic activity decays follows a power slope as a function of $R_o$ \citep[see figure~3 in][]{wrightetal2018}. For the sample of 51 K dwarf rotators with X-ray detection in the Hyades, the median $R_o$ = $0.46_{-0.08}^{+0.06}$ (N\'u\~nez et al, in prep.), which suggests that the lower levels of X-ray emission from K2-136 relative to its fellow Hyades K dwarfs can be explained by its slower rotation rate.

In conclusion, K2-136 does not appear unusually active in either the NUV or the X-ray relative to its fellow Hyades K dwarfs. 

\begin{table}[t]
\begin{center}
\caption{X-ray Spectral Fit Parameters of K2-136 \label{xrays_table}}
\begin{tabular}{llc}
\tableline
\tableline
Parameter\footnote[1]{All flux values are in the 0.1$-$2.4 keV energy range.} & Value & Unit \\
\tableline
\\
Degrees of Freedom & 33 &  \\
Reduced $\chi^2$ & 0.91 & \\
Plasma Temperature & 0.65$\pm$0.07 & keV \\
Plasma Metal Abundance & 0.06$\pm$0.02 & Solar \\
pn Energy Flux & 3.12$\pm$0.68 & 10$^{-14}$ erg s$^{-1}$ cm$^{-2}$ \\
MOS1 Energy Flux & 3.01$\pm$0.93 & 10$^{-14}$ erg s$^{-1}$ cm$^{-2}$ \\
MOS2 Energy Flux & 3.02$\pm$0.88 & 10$^{-14}$ erg s$^{-1}$ cm$^{-2}$ \\
EPIC\footnote[2]{The error-weighted average of the pn and MOS cameras.} Energy Flux & 3.06$\pm$0.46 & 10$^{-14}$ erg s$^{-1}$ cm$^{-2}$ \\

\\
\tableline
\end{tabular}
\end{center}
\end{table}

\subsection{Considering the Nearby Stellar Companion} \label{trend_analysis}

As discussed in Section~\ref{binarity}, prior observations of this system revealed a likely bound M7/8V companion with a projected separation of approximately 40 AU. The presence of a nearby stellar companion can easily cause a trend in RV observations. We wanted to estimate the range of possible trend amplitudes for our system using some reasonable assumptions. Using the distance of $58.752^{+0.061}_{-0.072}$ pc from \citet{bailer-jonesetal2021} and the projected angular separation of $0.730 \pm 0.030$'' from adding in quadrature the R.A. and Dec separation components in \citet{ciardietal2018}, we found a projected separation of $42.9 \pm 1.7$ AU. We considered the possibility of additional radial separation by folding in a uniform distribution on radial separation between 0 and twice the median projected separation to estimate an overall separation (this broad range was chosen to include radial separations of approximately the same scale as the projected separation). As an approximation, we treat this overall separation as the semi-major axis.

Next, the stellar companion was reported in \citet{ciardietal2018} to have a spectral type consistent with M7/8 (we were unable to find uncertainties associated with this result, but nearby spectral types were never mentioned). Taking a conservative approach, we assumed a stellar companion mass between $1$ M$_{\mathrm{Jup}}$ (for a low-mass brown dwarf) and $0.2$ M$_\odot$ (for a mid to late M dwarf).

Then, we combined our separation distribution and stellar mass distributions (primary and companion) via Kepler's Third Law to get a broad orbital period estimate of $520^{+320}_{-180}$ years. Including a wide range of eccentricities (Unif(0,0.9)), we estimated an RV semi-amplitude of $490^{+340}_{-320}$ m s$^{-1}$. On the timescale of our observations, this centuries-long sinusoidal signal would manifest as a linear trend, with a maximum (absolute) slope of $5.3^{+7.5}_{-3.6}$ m s$^{-1}$ yr$^{-1}$. This is a very rough estimate with many assumptions, but it serves to demonstrate that a drift of only a few m s$^{-1}$ each year or less is very reasonable. Indeed, from our model of the RV data we found an RV trend of $0.1 \pm 2.5$ m s$^{-1}$ yr$^{-1}$, highly consistent with both a zero trend as well as our estimate calculated here.

\section{Results and Discussion}\label{results_discussion}

The results of our stellar and planet analyses are listed in Tables~\ref{stellar_table}, ~\ref{rv_results}, and ~\ref{transit_results}. Phase plots of all three planets can be seen in Fig.~\ref{phase_plot}. After conducting our analysis and tests, we find that K2-136c has a mass of $18.1^{+1.9}_{-1.8}$ M$_\oplus$ and a radius of $3.00 \pm 0.13$ R$_\oplus$. This radius is consistent with and slightly larger than the value estimated in M18 ($2.91^{+0.11}_{-0.10}$ R$_\oplus$). This is because we find a stellar radius value slightly larger than M18 (by about $3\%$).

Using planet mass and radius we find K2-136c has a density of $3.69^{+0.67}_{-0.56}$ g cm$^{-3}$ (or $0.67^{+0.12}_{-0.10}$ $\rho_\oplus$). For comparison, Neptune\footnote[2]{https://nssdc.gsfc.nasa.gov/planetary/factsheet/neptunefact.html \label{neptuneref}} is roughly similar in mass ($17.15$ M$_\oplus$) but larger in radius ($3.883$ R$_\oplus$); as a result, K2-136c is more than twice as dense as Neptune ($2.25^{+0.41}_{-0.34}$ $\rho_{\neptune}$). Similarly, Uranus\footnote[3]{https://nssdc.gsfc.nasa.gov/planetary/factsheet/uranusfact.html \label{uranusref}} is slightly less massive ($14.54$ M$_\oplus$) but still larger in radius ($4.007$ R$_\oplus$); thus, K2-136c is nearly three times as dense as Uranus ($2.90^{+0.52}_{-0.44}$ $\rho_{\uranus}$). This is visually apparent in Fig.~\ref{mr_plot}, which shows K2-136c almost perfectly on the 100\% H$_2$O composition line. It is important to remember that mass and radius alone do not fully constrain a planet's composition. Although K2-136c may have a density similar to that of a large ball of water (an unrealistic reference composition), it is also consistent with a gaseous sub-Neptune with a massive core or metal-rich atmosphere.

Unfortunately, it is very difficult to determine compositional properties of a planet without atmospheric characterization, especially sub-Neptunes (since there are no analogs in our own Solar System). On the one hand, sub-Neptunes may include ocean worlds with H$_2$O abundance fractions not seen in our Solar System \citep{mousisetal2020}. And indeed, water vapor has already likely been detected in the atmosphere of the sub-Neptune exoplanet K2-18b \citep{bennekeetal2019b}. On the other hand, sub-Neptunes like K2-136c may instead be composed of a rocky, Earth-like core composition, very little water, and an atmosphere close to solar metallicity and thus primarily hydrogen and helium \citep{vaneylenetal2018,bennekeetal2019a}.

As a valuable point of comparison, there are three confirmed planets that share a similar mass and radius to K2-136c to within $10\%$: Kepler-276c, Kepler-276d \citep{xieetal2014}, and TOI-824b \citep{burtetal2020}. The masses of the planets in the Kepler-276 system were measured via transit timing variations (TTVs) in a TTV catalog paper. Unfortunately, because they were characterized alongside so many other systems, there is no discussion regarding the formation or composition of those two specific planets. TOI-824b, however, was characterized in a standalone paper that investigated the nature of the planet thoroughly. Unlike K2-136c, TOI-824b is near the hot-Neptune desert \citep{mazehetal2016} with a very short orbital period ($1.393$ d). Despite its proximity to its host star, TOI-824b still retains a H$_2$-He atmosphere, which the authors estimate has a mass fraction of $\geq 2.8\%$. They hypothesize that the larger than average mass of the planet (compared to planets of a similar radius) helps the planet retain its atmosphere. K2-136c, with a similar mass and radius and a lower insolation flux, would therefore be able to retain a H$_2$-He atmosphere even more easily.

We can go further and form a picture of a reasonable composition for K2-136c. We may assume the planet has a rocky core surrounded by a gaseous H$_2$-He envelope. Going a step further, we may also assume the rocky core is similar to that of Earth, namely a core-mass fraction (CMF) of $0.325$, i.e. a rock-iron composition of $32.5\%$ Fe and $67.5\%$ MgSiO$_3$ \citep{seageretal2007}. Following the theoretical models of \citet{howeetal2014}, we find that with an Earth-like rocky core and a H$_2$-He envelope, the measured mass and radius of K2-136c are most consistent with a H$_2$-He mass fraction of $\sim5\%$.

\begin{figure*}[ht!]
\epsscale{1.15}
  \begin{center}
      \leavevmode
\plotone{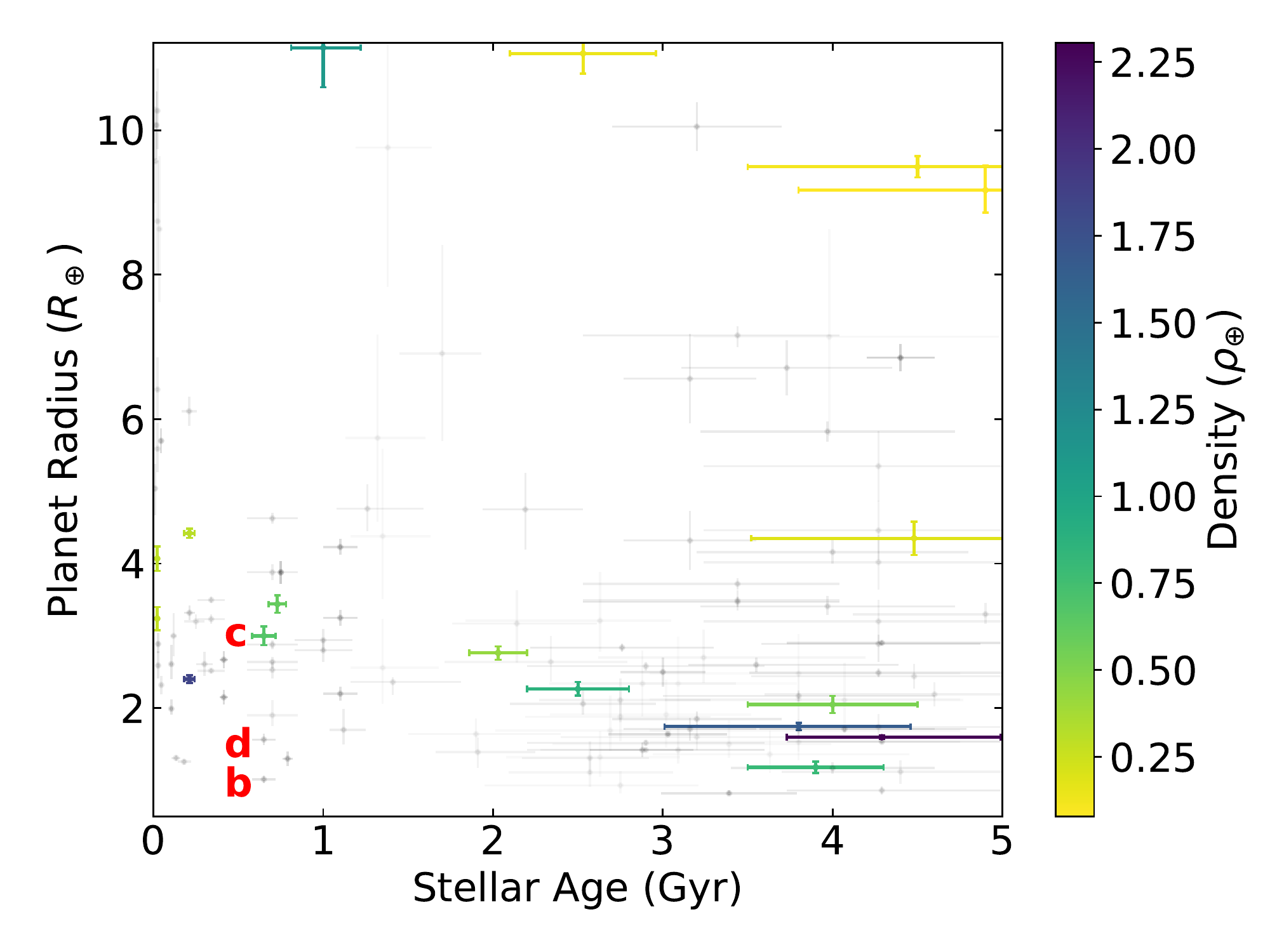}
\caption{Age-radius diagram for all planets smaller than Jupiter, younger than 5 Gyr, and with radius and age uncertainties both smaller than $25\%$. Data point color corresponds to planet density for planets with mass uncertainties smaller than $25\%$. Data collected from the NASA Exoplanet Archive (accessed 2023 Mar 22; \citealt{https://doi.org/10.26133/nea12}). K2-136b, c, and d are labeled in red to the left of the planet symbol. There are only four planets in this figure with a stellar age younger than K2-136 and a mass measurement better than $25\%$: AU Mic b and c (stellar age = $22 \pm 3$ Myr; \citealt{mamajekandbell2014}) and Kepler-411 b and c (stellar age = $212 \pm 31$ Myr; \citealt{sunetal2019}). The only other plotted planet $< 1$ Gyr with a mass measurement is the open cluster planet K2-25b \citep{stefanssonetal2020}, which is slightly larger than K2-136c.} \label{age_radius_plot}
\end{center}
\end{figure*}

With this rough envelope mass fraction estimate and the measured mass and radius of K2-136c, we wanted to investigate the potential for past or ongoing atmospheric mass loss. After consulting the theoretical models presented in \citet{lopezetal2012}, \citet{lopezandfortney2013}, \citet{lopezandfortney2014}, and \citet{jinetal2014}, we concluded that if there is any historical or contemporary mass loss for K2-136c, it is minimal: likely somewhere between $0-10\%$ loss of the H$_2$-He envelope across the entire lifespan of the planet.

As suggested in \citet{mannetal2016a}, young planets may be puffier than older planets due to an early-age atmospheric mass loss phase. And yet, K2-136c is not particularly puffy, in fact being notably denser than Uranus or Neptune. However, this planet may indeed have an extended atmosphere but also a lower atmospheric mass fraction than Uranus, Neptune, and other lower-density planets. In other words, as this system ages, the atmosphere of K2-136c may settle to some extent, reducing the planet radius and increasing planet density. Without atmospheric characterization, further insights into the planet's composition are very limited.

Bayesian model comparison proved that we could not conclusively detect K2-136b or K2-136d in our data set (see Fig.~\ref{phase_plot}). Even so, we also conduct similar analyses for K2-136b and K2-136d in order to report upper mass limits and corresponding upper density limits. With 95\% confidence, K2-136b is no denser than $24$ g cm$^{-3}$ (a largely unhelpful limit, since that would be much more dense than pure iron) and K2-136d is no denser than $4.3$ g cm$^{-3}$, corresponding to semi-amplitudes of $1.7$ m s$^{-1}$ and $0.78$ m s$^{-1}$, respectively. Referring to Fig.~\ref{mr_plot}, we can see that unlike the middle planet K2-136c, the other two planets K2-136b and K2-136d could have a wide variety of densities and compositions. K2-136d could range from a low-density gas planet to Earth composition. K2-136b has an even wider array of possible compositions and theoretically could range from very gaseous to pure iron.

The RV signals of K2-136b and K2-136d may be detectable with more data from next generation spectrographs. K2-136d would be particularly interesting, since its radius places it near the planet radius gap \citep{fultonetal2017}. As for K2-136b, the peak of the planet mass posterior distribution is already non-zero, and a marginal detection may already be noted at the $2.0\sigma$ level ($2.4 \pm 1.2$ M$_\oplus$), suggesting a firm planet mass measurement may be within reach with further observations.

To check this, we followed the mass-radius relationship laid out in \citet{wolfgangetal2016} and found predicted masses for K2-136b and K2-136d to be $1.17^{+0.79}_{-0.72}$ M$_\oplus$ and $4.1^{+1.9}_{-1.8}$ M$_\oplus$, respectively. These correspond to densities of $7.7^{+3.7}_{-4.7}$ g cm$^{-3}$ (i.e. $1.40^{+0.66}_{-0.85}$ $\rho_\oplus$) and $8.4^{+4.1}_{-3.7}$ g cm$^{-3}$ (i.e. $1.52^{+0.74}_{-0.66}$ $\rho_\oplus$), respectively. We note that these mass and density estimates do not make use of the upper limits determined in this paper. By folding in our stellar mass, orbital period, and eccentricity posteriors as well as the orbital inclinations determined in M18, we found the estimated masses of K2-136b and K2-136d correspond to semi-amplitudes of $0.5 \pm 0.3$ m s$^{-1}$ and $1.1 \pm 0.5$ m s$^{-1}$, respectively. The current RV upper limit on K2-136b ($1.7$ m s$^{-1}$) is much larger than the estimated semi-amplitude and therefore fully consistent. As for K2-136d, we acknowledge that the estimated semi-amplitude is smaller than the upper limit ($0.80$ m s$^{-1}$), which perhaps suggests K2-136d has a density on the lower end of the range predicted from the \citet{wolfgangetal2016} relationship.

There are very few young and small exoplanets that also have measured masses. As can be seen in Fig.~\ref{age_radius_plot}, the vast majority of young exoplanets do not have a firm mass measurement. There are some young planets with both robust radius measurements and notable upper mass limits, such as the low-density planet TS Duc A b \citep{benattietal2021}, which can be of interest for follow-up study and comparison. However, according to the NASA Exoplanet Archive (accessed 2023 Mar 12; \citealt{https://doi.org/10.26133/nea12}), there are only 13 known planets, excluding K2-136c, with $R_p < 4$ R$_\oplus$, a host star age $< 1$ Gyr, and a mass measurement (not an upper limit): HD 18599b \citep{desideraetal2022}, HD 73583b and c \citep{barraganetal2022}; K2-25b \citep{stefanssonetal2020}; L 98-59b, c, and d \citep{demangeonetal2021}; Kepler-411b and Kepler-411d \citep{sunetal2019}, Kepler-462b \citep{masudaetal2020}, Kepler-289b and Kepler-289d \citep{schmittetal2014}, and K2-100b \citep{barraganetal2019}.

K2-136c is now the smallest exoplanet in an open cluster to have a mass measurement. It is also one of the youngest exoplanets to ever have a mass measurement. The only planets with firm age, radius, and mass measurements ($<25\%$ uncertainties) known to be younger are AU Mic b and c \citep{kleinetal2021} as well as Kepler-411b and d \citep{sunetal2019}, as can be seen in Fig.~\ref{age_radius_plot}. In general, measuring the masses of young planets like K2-136c provides an interesting window into the early childhood of planetary systems, allowing us to probe how planet masses and compositions evolve over time.

\begin{figure}
    \adjustimage{width=.48\textwidth,left}{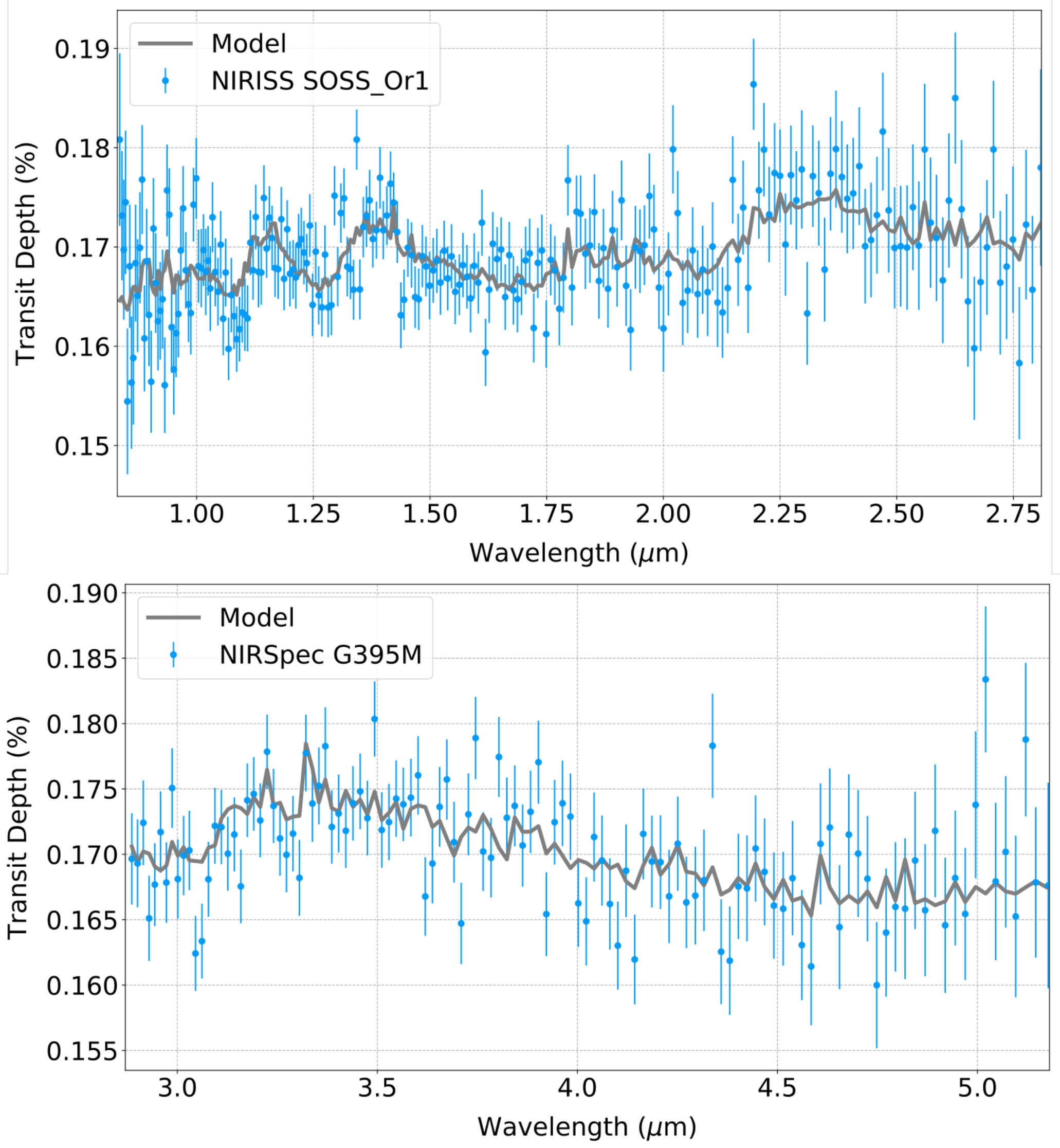}
    \caption{Simulation of \emph{JWST} transmission spectra for K2-136c using \texttt{JET} \citep{fortenbachanddressing2020}. The top and bottom panels correspond to the NIRISS SOSS-Or1 (0.81-2.81 $\mu$m) instrument, and the NIRSpec G395M (2.87-5.18 $\mu$m) instrument, respectively. The gray line in both panels is the modelled atmospheric spectrum assuming low metallicity (5x solar) and no clouds, while the blue data points are the simulated instrument spectra for one observed transit with NIRISS SOSS, and two observed transits with NIRSpec G395M, including the effects of photon noise, and instrument systematics.} \label{jet}
\end{figure}

\subsection{Atmospheric Characterization Prospects}

To explore the suitability of the K2-136 system planets for atmospheric characterization, we calculated the transmission spectroscopy metric (TSM) defined in \citet{kemptonetal2018}. K2-136c has a TSM of $32.7^{+4.8}_{-4.1}$, which is well below the recommended TSM of $90$ for $R_p > 1.5$ R$_\oplus$. Because K2-136b and K2-136d have unconstrained masses, we followed \citet{zengetal2016} and assumed an Earth-like CMF of $0.325$ in order to predict planet masses of $0.85^{+0.27}_{-0.21}$ M$_\oplus$ and $3.35^{+1.08}_{-0.84}$ M$_\oplus$, respectively. For K2-136b, this yields a TSM of $4.20^{+0.42}_{-0.37}$, well below the recommended TSM of $10$ for $R_p < 1.5$ R$_\oplus$. As for K2-136d, its radius of $R_d = 1.565 \pm 0.077$ R$_\oplus$ is very near $1.5$ R$_\oplus$, where the TSM metric includes a scale factor that jumps dramatically, thus creating a TSM bimodal distribution. Thus, for $R_d < 1.5$ R$_\oplus$ we find a TSM of $2.36^{+0.15}_{-0.13}$ and for $R_d > 1.5$ R$_\oplus$ we find a TSM of $13.6^{+1.0}_{-1.1}$. In their respective radius ranges, these values are both well below the recommended TSM, so K2-136d is also probably not a good target for atmospheric characterization.

We also calculated the emission spectroscopy metric (ESM) for K2-136b and K2-136d as defined in \citet{kemptonetal2018} (the metric applies only to ``terrestrial'' planets with $R_p < 1.5$ R$_\oplus$, excluding K2-136c). We find K2-136b and K2-136d have ESM metrics of $0.520^{+0.049}_{-0.047}$ and $0.342^{+0.043}_{-0.041}$, respectively, both below the recommended ESM of $7.5$ or higher. Therefore, these planets do not appear to be particularly attractive targets for emission spectroscopy or phase curve detection. 

The TSM analysis of K2-136c is not very favorable for atmospheric characterization, but we decided to conduct a more thorough transmission spectroscopy analysis. We used the \texttt{JET} tool \citep{fortenbachanddressing2020} to model atmospheric spectra and to simulate the performance of the \emph{JWST} instruments for certain atmospheric scenarios. We opted for the broad wavelength coverage of combining NIRISS SOSS Order 1 (0.81-2.81 $\mu$m) with NIRSpec G395M (2.87-5.18 $\mu$m), as recommended by \citet{batalhaandline2017} to maximize the spectral information content. A single instrument, the NIRSpec Prism, can also cover this wavelength range, but brightness limits preclude its use here. We assumed pessimistic pre-launch instrumental noise values \citep{rigbyetal2022}, so a future \emph{JWST} program for atmospheric characterization should outperform our conservative expectations.

The \texttt{JET} tool found that for an optimistic, cloudless, low-metallicity atmosphere (5x solar) we can meet a $\Delta$BIC detection threshold of 10 (corresponding to a $\sim3.6\sigma$ detection of the atmosphere compared to a flat line) with 5 free retrieval parameters (i.e., recon level) with only one transit for NIRISS SOSS and two transits for NIRSpec G395M. For a less optimistic, higher metallicity atmosphere (100x solar) with clouds at 100 mbar, we can meet the same detection threshold with two transits for NIRISS SOSS and five transits for NIRSpec G395M.

With 10 free retrieval parameters (a more typical number), and with the optimistic atmosphere, we can meet a $\Delta$BIC detection threshold of 10 with only one transit for NIRISS SOSS and three transits for NIRSpec G395M. For the less optimistic atmosphere, we can meet the detection threshold with three transits for NIRISS SOSS, but we show no detection for NIRSpec G395M for up to 50 transits considered. This analysis makes the conservative assumption that the instrument noise floor is not reduced by co-adding transits.

The resulting spectra for the low metallicity, cloudless, case are shown in Fig.~\ref{jet}. It seems that atmospheric characterization of K2-136c may be within reach (assuming a relatively low mean molecular weight/low metallicity atmosphere, and low cloud level), but could require a more significant investment of \emph{JWST} resources if the actual atmospheric properties are less favorable. 

It should be noted that given the on-sky position of K2-136, the ability to observe the system with JWST will be limited due to aperture position angle constraints. In addition, the very close ($\sim0.7$'') stellar companion may cause contamination of spectra from both instruments. This is a common issue for NIRISS since it is slit-less, but NIRSpec can usually isolate the primary target with its $1.6$'' square aperture. For K2-136 the companion star is well inside this aperture boundary and will likely create some contamination. The companion is significantly fainter than the host star (J magnitude of $14.1$ vs $9.1$), which should mitigate the impact to a degree. It should also be possible to reduce the companion M-star’s spectral contamination effect in post-processing.

The most enticing feature of K2-136 is the young age of the system; it could be argued that despite the potential difficulty in observing the planets' atmospheres, the rewards outweigh the risks for the chance to better understand the atmospheres of very young, relatively small planets. Observations of this system could help us construct a picture of the environment and evolution of young, low-mass planets.

\section{Summary and Conclusions} \label{conclusion}

In this paper, we analyzed K2-136, a young system in the Hyades open cluster. The star is a K dwarf with $M_* = 0.742^{+0.039}_{-0.038}$ M$_\odot$ and $R_* = 0.677 \pm 0.027$ R$_\odot$. It hosts three known, transiting planets with periods of $8.0$, $17.3$, $25.6$ days, and radii of $1.014\pm0.050$ R$_\oplus$, $3.00\pm0.13$ R$_\oplus$, and $1.565\pm0.077$ R$_\oplus$. We gathered RV observations with the TNG HARPS-N spectrograph and ESPRESSO VLT spectrograph in order to measure the masses of the three planets. We find that K2-136c, a sub-Neptune and the middle planet of the system, has a mass of $18.0^{+1.7}_{-1.6}$ M$_\oplus$. This corresponds to a density of $3.69^{+0.67}_{-0.56}$ g cm$^{-3}$ (or $0.67^{+0.12}_{-0.10}$ $\rho_\oplus$). K2-136c is thus similar in mass to Neptune and Uranus but more than twice as dense as Neptune and nearly three times as dense as Uranus. K2-136c has a density consistent with an ocean world; a rocky, Earth-like core with solar metallicity atmosphere; and many other compositions. However, assuming an Earth-like rocky core and a H$_2$-He envelope yields a H$_2$-He mass fraction of $\sim5\%$. K2-136b and K2-136d have RV signals too small to detect with our data set, but we have placed upper mass limits with 95\% confidence of $4.3$ and $3.0$ M$_\oplus$, respectively. Atmospheric characterization of K2-136c (or its siblings, if a firm mass measurement can be made), would be difficult but not necessarily unfeasible, and is the most practical way to narrow the compositional parameter space for these planets.

K2-136c is the smallest planet in an open cluster to have a mass measurement, and one of the youngest planets found to date smaller than Neptune. There are very few young planets with precise mass measurements, and even fewer as small as K2-136c. As a result, this system provides an important view of planet composition and evolution at ages that are relatively unexplored.

\acknowledgments

We are grateful to Eugene Chiang for useful feedback and advice regarding planetary formation and composition in this system.

A.W.M. is supported by the NSF Graduate Research Fellowship grant no. DGE 1752814. C.D.D. acknowledges support from the NASA K2 Guest Observer program through grant 80NSSC19K0099, the Hellman Family Faculty Fellowship, the Alfred P. Sloan Foundation (Grant FG-2019-11662), and the David \& Lucile Packard Foundation (Grant 2019-69648). 

This work has been partially supported by the National Aeronautics and Space Administration under grant No. NNX17AB59G.

The financial contribution from the agreement ASI-INAF n.2018-16-HH.0 is gratefully acknowledged.

Support for this work was provided by NASA through grants number HST-GO-15090.001-A and HST-GO-15091.001-A from the Space Telescope Science Institute, which is operated by AURA, Inc., under NASA contract NAS 5-26555.

This paper includes data collected by the \Kepler\ mission. Funding for the \Kepler\ mission was provided by the NASA Science Mission directorate.

Based on observations made with the Italian {\it Telescopio Nazionale Galileo} (TNG) operated by the {\it Fundaci\'on Galileo Galilei} (FGG) of the {\it Istituto Nazionale di Astrofisica} (INAF) at the {\it Observatorio del Roque de los Muchachos} (La Palma, Canary Islands, Spain).

LBo, GPi, IPa, VNa, and GLa acknowledge the funding support from Italian Space Agency (ASI) regulated by ``Accordo ASI-INAF n. 2013-016-R.0 del 9 luglio 2013 e integrazione del 9 luglio 2015 CHEOPS Fasi A/B/C''.

FL gratefully acknowledges a scholarship from the Fondation Zd\u{e}nek et Michaela Bakala.

A.Mo. acknowledges support from the senior Kavli Institute Fellowships.

Some of the data presented in this paper was obtained from the Mikulski Archive for Space Telescopes (MAST). STScI is operated by the Association of Universities for Research in Astronomy, Inc., under NASA contract NAS5--26555. Support for MAST for non--HST data is provided by the NASA Office of Space Science via grant NNX13AC07G and by other grants and contracts.

This work has made use of data from the European Space Agency (ESA) mission {\it Gaia} (\url{https://www.cosmos.esa.int/gaia}), processed by the {\it Gaia} Data Processing and Analysis Consortium (DPAC, \url{https://www.cosmos.esa.int/web/gaia/dpac/consortium}). Funding for the DPAC has been provided by national institutions, in particular the institutions participating in the {\it Gaia} Multilateral Agreement.

This paper includes data collected with the TESS mission, obtained from the MAST data archive at the Space Telescope Science Institute (STScI). Funding for the TESS mission is provided by the NASA Explorer Program. STScI is operated by the Association of Universities for Research in Astronomy, Inc., under NASA contract NAS 5–26555.

Facilities: \facility{Kepler, TESS, Gaia, HST, VLT, TNG, XMM-Newton, NASA Exoplanet Archive, ADS, MAST}

Software: ARESv2 \citep{sousaetal2015}, ATLAS \citep{kurucz1993}, \texttt{BATMAN} \citep{kreidberg2015}, \texttt{isochrones} \citep{morton2015a}, \texttt{JET} \citep{fortenbachanddressing2020}, \texttt{lightkurve} \citep{barentsenetal2019}, MOOG \citep{sneden1973}, \texttt{MultiNest} \citep{ferozetal2009,ferozetal2013}, \texttt{RadVel} \citep{fultonetal2018a}, \texttt{SPOCK} \citep{tamayoetal2020}

\bibliographystyle{apj}
\bibliography{refs}

\clearpage

\end{document}